\documentclass[12pt]{article}
\usepackage{amssymb,amsmath,amsthm}
\usepackage[mathscr]{eucal}
\usepackage{array}
\usepackage[dvips]{graphicx}  
\makeatother
\newcommand{\h}{\hspace*{1pt}}
\providecommand{\sh}{\mathop{\rm sh}\nolimits}
\providecommand{\sign}{\mathop{\rm sign}\nolimits}
\providecommand{\mod}{\mathop{\rm mod}\nolimits}

\newcommand{\pa}{\partial}
\newcommand{\op}{\operatorname}

\def\CH{\mathcal{ H}}

\makeatletter
\@addtoreset{equation}{section}

\newtheorem{theorem}{Theorem}

\newtheorem{lemma}{Lemma}
\newtheorem{rem}{Remark}
\newtheorem*{cor}{Corollary}

\newtheorem{exam}{Example}
\newtheorem{definition}{Definition}

\providecommand{\sh}{\mathop{\rm sh}\nolimits}
\providecommand{\sign}{\mathop{\rm sign}\nolimits}
\providecommand{\mod}{\mathop{\rm mod}\nolimits}

\begin{document}
\title
{Generalized Maslov canonical operator and tsunami asymptotics over nonuniform bottom. I}
 \sloppy
\author{
     Sergey Dobrokhotov\footnote{
Institute for Problems in Mechanics, RAS,
Moscow;
E-mail: dobr{@}ipmnet.ru}
\and
Sergey Sekerzh-Zenkovich\footnote{
Institute for Problems in Mechanics, RAS,
Moscow;
E-mail: seker{@}ipmnet.ru}
\and
Brunello Tirozzi\footnote{Department of Physics,
University "La Sapienza", Rome;
E-mail: tirozzi{@}krishna.phys.uni-roma1.it
}
\and 
Timur Tudorovskiy \footnote{
Institute for Problems in Mechanics, RAS,
Moscow;
E-mail: timtudor{@}ipmnet.ru}} 
\maketitle

\sloppy
\sloppy \abstract {We suggest a new asymptotic
 representation for the solutions to the 2-D wave e
 quation with variable velocity with localized initial data. 
 This representation is a generalization of the 
 Maslov canonical operator and gives the formulas 
 for the relationship between initial localized 
 perturbations and  wave profiles near the wave 
 fronts including the neighborhood of backtracking 
 (focal or turning) and selfintersection points. 
 We apply these formulas to the problem of a 
 propagation of tsunami waves in the frame of 
 so-called piston model. Finally we suggest the  
 fast asymptotically-numerical algorithm for simulation 
 of tsunami wave over nonuniform bottom. In this first part we
 present the final formulas and some geometrical construction.
 The  proofs concerning analytical calculations   
 will be done in the second part.}       

{\section{Main equations and a simple example: the wave field in the case of constant bottom}

\subsection {Some notation} Let us introduce the notations used in this paper. A two
dimensional vector can be written with capital or small letters
$X=(X_1,X_2)$ or $x=(x_1, x_2)$. The vector can be
written also as a column vector $\begin{pmatrix} X_1 \\
X_2\end{pmatrix} $. Two dimensional vectors $X$ and $Y$ can form a
column vector $\begin{pmatrix} X
\\ Y \end{pmatrix} $ with four rows. The real scalar product between
two vectors $X$ and $Y$, with real components, is indicated by $ <
X, Y>$, the complex scalar product among bi-dimensional vectors
$Z$, $W$, with complex components, is written as $<Z, W>_c$, the
two by two matrix generated by two bi-dimensional vectors $X$, $Y$
is written as $(X,Y)$ where in the first column there are the
components of the vector $X$ and in the second column those of the
vector $Y$; the transposed matrix of $C$ is denoted by $^t C$.

\subsection{Main equations}
Let us remind the  statements of problems used in  tsunami 
wave  problems (\cite{Voit}-\cite{Ivanov})
 as well as in general linear water wave theory 
 (\cite{Stoker}-\cite{KurkPeli}.

Let us assume that the bottom of the basin is moving $H=H_0(x)+H_1(x,t)$.
We assume also that the
perturbation $H_1(x,t)$ is small with respect to  $H_0$ \break $|H_1|<<H_0(x)$,
and that $H_1$ is localized in a neighborhood
of some given point $x_0$. If $L$ is the dimension of the region where
the wave phenomena is studied, and $l$ is the dimension of
the perturbed region, then our hypothesis implies that $l<<L$.
Another assumption is that the bottom "changes slowly", i.e. that
$\nabla H_0\sim \mu$, where $\mu$ is some small (''adiabatic parameter").
We discuss below its meaning.
 Introducing the scaled variables $x'=\frac {x}{L}$, then
 $H=H_0(x')+\chi H_1(\frac {x'}{\mu},t)$, where $\mu=\frac l L<<1$.

The equation for the velocity potential $\Phi $ in the water $-H\leq z\leq \eta$,
where $\eta (x,t)$ is the sea
elevation in the linear approximation, has the form, in dimensional variables,:
\begin{equation}
\Delta\Phi=0,
\label{1.1}
\end{equation}
\begin{equation}
\eta_t-\frac{\partial\Phi }{\partial z}|_{z=0}=0,
\label{1.2}
\end{equation}
\begin{equation}
\Phi _t+g \eta|_{z=0}=0,
\label{1.3}
\end{equation}
\begin{equation}
\frac{\partial\Phi }{\partial n}\equiv \frac{\partial\Phi
}{\partial z}+ <\nabla H,\nabla\Phi>=v(x,t)|_{z=-H}. \label{1.4}
\end{equation}
where $v(x,t)$ is the normal component of the velocity of the motion of the bottom in the point $x$.
%%%%%%%%%%%%%%%%%%%%%%%%%%%%%%%%%%%%
%%%%%%%%risynok
The velocity $v$ can be expressed by means of the derivative
$\frac{\partial H _1}{\partial t}$ by : $\frac{\partial H
_1}{\partial t}/\sqrt {(\nabla H)^2+1}$, since $V$ is the
projection of the velocity on the vector $\frac {1} {\sqrt
{(\nabla H)^2+1}}^t(\nabla H,1)$ normal to the surface $z=-H.$ If
we consider $\nabla H_0$ to be small (because of the slow
variation of the bottom relief), and that also $\nabla H_1$ is
small (because of the small amplitude $H_1$), then we have
$$v=\frac{\partial H_1}{\partial t}.$$

\subsection{The solution in the form of the Fourier transform}

Let us begin considering the system (\ref{1.1})-(\ref{1.4}) in the
case of constant bottom.  In this case the velocity potential and
its derivatives are zero for $t=0$. We make the Fourier transform
of the system (\ref{1.1})-(\ref{1.4}) with respect to the
variables $x_1,x_2$. The dual variables will be denoted with
$p_1,p_2$ and the Fourier transform of the corresponding function
will be considered as a  ``wave''. Then (\ref{1.1})-(\ref{1.4})
gets the form
\begin{equation}
\widetilde\Phi_{zz}-p^2\widetilde\Phi=0,
\label{4.1}
\end{equation}
\begin{equation}
\widetilde\eta_t-\frac{\partial\widetilde\Phi }{\partial z}|_{z=0}=0,
\label{4.2}
\end{equation}
\begin{equation}
(\widetilde\Phi _t+g \widetilde\eta)|_{z=0}=0,
\label{4.3}
\end{equation}
\begin{equation}
\frac{\partial\widetilde\Phi }{\partial z}|_{z=-H}=\tilde v\equiv\frac{\pa \tilde H_1}{\pa t}.
\label{4.4}
\end{equation}
Solving  (\ref{4.1})-(\ref{4.4}), we find
\begin{equation}
\widetilde\Phi =\frac{\cosh((z+H)|p|)}{\cosh H|p|}\tilde \varphi+
\frac{\sh (z|p|)}{|p|\cosh(H|p|)}\frac{\partial\widetilde{H_1}}{\partial t}
\label{4.5}
\end{equation}
and
\begin{equation}
\widetilde\Phi_z|_{z=0}=|p|\tanh(H|p|)\tilde \varphi+\frac{1}{\cosh{H|p|}}
\frac{\partial\widetilde{H_1}}{\partial t}
\label{4.6}
\end{equation}
Thus the equations (\ref{4.2})-(\ref{4.3}) take the form
\begin{gather}
\frac{\partial\tilde{\eta}}{\partial t}-|p|\tanh(H|p|)\tilde\varphi-
\frac{1}{\cosh(H|p|)} \frac{\partial\widetilde{H_1}}{\partial t}=0\nonumber\\
\frac{\partial\tilde{\varphi}}{\partial t}+g\tilde {\eta}=0
\label{4.7}
\end{gather}
Where $\tilde \varphi= \widetilde\Phi _t|_{z=0}$, and we have the initial conditions $t=0\qquad$
\begin{equation}
\tilde{\varphi}|_{t=0}=0,\qquad\tilde{\varphi}_t|_{t=0}=0 \Longleftrightarrow \tilde{\eta}|_{t=0}=0.
\label{4.8}
\end{equation}
These conditions define the so called Cauchy-Poisson problem for
the system (\ref{4.7}). They are compatible with the perturbation
of the bottom only if we suppose that the earthquake starts at at
a time different from zero. So we assume that the bottom has an
"instantaneous" $\quad$ movement at a small time  $t=\varepsilon$:
\begin{equation}\label{H1}
H_1(x,t)=\theta(t-\varepsilon)V(x),
\end{equation}
then we send $\varepsilon$
to zero at the end of the calculation; the smooth function $V(x)$ decays rapidly at infinity.

Differentiating  the first equation in (\ref{4.7}) with respect to $t$ and substituting $\frac{\partial \tilde {\varphi}}{\partial t}$ with $-g\tilde {\eta}$
we get the equation for $\tilde {\eta}$:
\begin{equation}
\tilde {\eta}_{tt}+\mathcal{L}\tilde {\eta}-
\frac{1}{\cosh(H|p|)} \frac{\partial^2\widetilde{H_1}}{\partial t^2}=0, \qquad \mathcal{L}=g|p|\tanh(H|p|).
\label{4.9}
\end{equation}
Differentiating the second equation of the system (\ref{4.7}) with respect to $t$ and substituting the derivative $\eta_t$
with the expression of the first equation and considering the condition that the source is active at the moment $t=\varepsilon>0$,
we get $\varphi_{tt}|_{t=0}=-g|p|\tanh(H|p|)\tilde\varphi |_{t=0}=0$ and the initial condition for  (\ref{4.9})
\begin{equation}
\eta_{t=0}=0\quad \eta_t|_{t=0}=0.
\label{4.10}
\end{equation}
It is easy to find the solution $\widetilde G$ of the homogeneous equation associated with (\ref{4.9}):
\begin{gather*}
\widetilde G_{tt}+\mathcal{L}(p,H)\widetilde G=0,\quad  \widetilde G|_{t=\tau}=0,\quad \widetilde G_t|_{t=\tau}=1,\\ \widetilde G(t,\tau,p)=\frac{e^{i\sqrt \mathcal{L}(t-\tau)}-e^{-i\sqrt \mathcal{L}(t-\tau)}}{2i\sqrt \mathcal{L}}=\frac{\sin\sqrt \mathcal{L}(t-\tau)}{\sqrt \mathcal{L}}.
\end{gather*}
In this way the solution of the non homogeneous equation (\ref{4.9}) is
$$
\tilde \eta=\int_0^t\widetilde G(t,\tau,p)\frac{1}{\cosh(H|p|)}\frac{\partial^2\widetilde H_1(\tau,p)}{\partial t^2}d\tau.
$$
The inverse Fourier transform of the function $\tilde \eta$ gives
the elevation of the free surface. Under our assumption of
instantaneous motion at time $\varepsilon$ we have
$\frac{\partial^2\widetilde H_1(\tau,p)}{\partial
t^2}=\delta'(t-\varepsilon)\widetilde V$ and so:
\begin{gather*}
\tilde \eta=\int_0^t\widetilde G(t,\tau,p)\frac{1}{\cosh(H|p|)}\frac{\partial^2\widetilde H_1(\tau,p)}{\partial t^2}d\tau
=\frac{\widetilde V}{\cosh H|p|}\int_0^t\frac{\sin\sqrt \mathcal{L}(t-\tau)}{\sqrt \mathcal{L}}\delta'(\tau-\varepsilon)d\tau=\\
%(\Theta_{\tau\tau}(t-\tau)=\delta'(\tau-\varepsilon))\\
-\frac{\widetilde V}{\cosh H|p|}\frac {\partial}{\partial\tau}\big(\frac{\sin\sqrt \mathcal{L}(t-\tau)}{\sqrt \mathcal{L}}\big)\big |_{\tau=\varepsilon}=
\frac{\widetilde V}{\cosh H|p|}\cos\sqrt \mathcal{L}(t-\varepsilon).
\end{gather*}
We send  now $\varepsilon$ to zero so we get the function
$\tilde \eta=\frac{\widetilde V}{\cosh(H|p|)}\cos\sqrt \mathcal{L}(t)$. It is evident that $ \tilde \eta $ is the solution of the equation (\ref{4.9})
with the following Cauchy conditions
 \begin{equation}
\tilde \eta|_{t=0}\equiv\frac {\widetilde V}{\cosh (H|p|)},\quad \tilde \eta'|_{t=0}=0.
\label{4.11}
\end{equation}

We shall discuss the meaning of such initial conditions for the function $\eta$ later.

\subsection{Solution of the Cauchy problem for constant bottom and instantaneous source}
%Рассмотрим сначала задачу \eqref{2.14} в случае постоянного дна (и мгновенного по времени источника). В этом случае  операторы $\widehat L$ и $\widehat M$ вычисляются точно.  Обозначим
%$\eta|_{t=0}=\eta_0( x)$. Применим к \eqref{2.14} преобразование Фурье, это дает
%\begin{equation*}\tilde \eta_{tt}+\mathcal{L}(p,H)\tilde \eta=0,\quad \tilde \eta|_{t=0}=\tilde \eta_0(p),\quad \tilde \eta'_t|_{t=0}=0.\end{equation*}
%Здесь $\mathcal{L}(p,H)=g|p|\tanh(|p|H),\quad$ $\tilde \eta_0(p)=\frac {1}{\cosh(|p|H)}\tilde V$, $\quad\tilde V$--преобразование Фурье функции $V$.

Let us study the solution $\eta$ corresponding to (\ref{4.11}). It
is not restrictive to assume that the center of the source is
located in the origin of the coordinates $x_0=0$ and that the
perturbation decays rapidly with the distance from the origin and
that it has a maximum in a small neighborhood of the origin. We
use also dimensionless variables:
$$
V=V(\frac {x}{l}),
$$
where  $l$ is the size of the shifted region and
%\begin{equation}
\begin{eqnarray*}
&\tilde{V}=\frac {1}{2\pi}\int V(\frac {\xi}{l})
e^{-ip\cdot\xi}d\xi=
%\frac{1}{2\pi}e^{-ip\cdot x_0}\int V(\frac {\xi}{l})
%e^{-ip\cdot (\xi)}d\xi=\\
\frac{l}{2\pi}\int V(y)e^{-i l <p, y>}dy= l\tilde{V}(pl),\\
&\tilde \eta_0(p)=\frac {l}{\cosh(|p|H)}\tilde V(pl),
\end{eqnarray*}
%\end{equation}

\noindent where we made the substitution $(y=\frac {\xi }{l},\quad
\xi=yl)$ and $\tilde V(p)$ is the usual Fourier transform of
the function $V(y)$. We assume that $V(y)$ is smooth function rapidly
decaying as $|y|\to \infty$.
%Решая эту задачу, получим
%\begin{equation*}\tilde \eta=\frac 1 2(e^{it\sqrt{\mathcal{L}(p,H)}}+ e^{-it\sqrt {\mathcal{L}(p,H)}})\tilde \eta_0(p),
%\end{equation*}и
%\begin{equation}\eta=,\label{2.1}\end{equation}
%где
%$${\tilde \eta}_0(p)=\frac {1}{2\pi}\int \eta_0(\xi)e^{-ip \cdot \xi}d\xi=\frac {1}{2\pi\cosh(|p|H)}\int \tilde V(\xi)e^{-ip \cdot \xi}d\xi$$

Then we can make the inverse Fourier transform:
$$\eta=\frac {l}{4\pi}\sum_{\pm}\int e^{\pm
it\sqrt{\mathcal{L}(p,H)}+i<p, x>} \tilde \eta_0(p)dp=\frac
{l}{4\pi}\Sigma _{\pm}\int e^{\pm it\sqrt{\mathcal{L}(p,H)}+i<p,
x>}\frac {1}{\cosh(|p|H)}\tilde V(pl)dp.
$$
Changing the variables $pl=p',p=p'/l$, we get
$$
\eta=\frac {1}{4\pi}\Sigma _{\pm}\int e^{\pm i\sqrt{\frac
{g|p|}{l}\tanh(|p|\frac{H}{l})}+i\frac{<p,x>}{l}}\frac{1}{\cosh(|p|\frac{H}{l})}\tilde
V(p)dp.
$$
In this way the problem is reduced to the computation of the asymptotic behavior of the integral.

We will study the asymptotic values for  $|x|>> l$. We change to polar coordinates $(\rho,\varphi)$ in the integral, where $\varphi$ is defined as the angle
among  $p$ and $x-x_0$. Thus \break $p=\rho\Theta(\varphi) \frac{x}{|x|}$, where   $\Theta(\varphi)$ is the two dimensional matrix defining the rotation
of an angle $\varphi$.
$$
\Theta(\varphi)=\begin{pmatrix}
\cos\varphi &-\sin \varphi\\\sin\varphi& \cos \varphi
\end{pmatrix}
$$
Then the last integral has the form
\begin{gather*}
\eta=\frac {1}{4\pi}\Sigma _{\pm}\int _0^\infty\rho d\rho \int _0^{2\pi}d\varphi \exp\big(\pm i t\sqrt{\frac {g\rho}{l}\tanh (\rho\frac{H}{l})}\big)
 \exp\big({i\frac{\rho|x|}{l}\cos \varphi}\big)
\frac{1}{\cosh(\rho\Theta\frac {x}{|x|}\frac{H}{l})}\tilde V(\rho\Theta\frac {x}{|x|}).
%\tilde{\eta}_0(\rho\Theta\frac {x}{|x|})dp
\end{gather*}
The internal integral can be computed using the method of
stationary phase. The phase has the form:
$\Phi=\frac{\rho|x|}{l}\cos \varphi$, the equation
$\frac{\partial\Phi}{\partial\varphi}=0$ gives
$\varphi=0,\varphi=\pi$; furthermore it is not possible to apply
the method of the stationary phase in the point  $\rho=0$.
Nevertheless one can take a sufficiently small neighborhood of the
saddle points of the variable $\varphi$ and show that,
\cite{borov,Borovikov, DobrZhKuz,berry}, it is smaller than
the contribution of the terms that we neglect. The result is:
\begin{gather*}
\rho\int _0^{2\pi}d\varphi e^{\frac{i\varrho|x|}{l}\cos \varphi}\frac{1}{\cosh(\rho\widehat{\Theta}\frac {x}{|x|}\frac{H}{l})}
\tilde V(\rho\widehat{\Theta}\frac{x}{|x|})\approx\\
\frac{\sqrt{2\pi}}{\cosh(\rho\frac {x}{|x|}\frac{H}{l})}(
e^{-i\pi/4}\sqrt
{\frac{l\varrho}{|x|}}e^{\frac{i\varrho|x|}{l}}\tilde V(\rho
\frac{x}{|x|})+e^{i\pi/4}\sqrt{\frac{l\varrho}{|x|}}
e^{\frac{-i\varrho|x|}{l}}\tilde V(-\rho\frac{x}{|x|}))%+O(\rho)
\end{gather*}
and
\begin{gather*}
\eta\approx\frac {1}{2\sqrt{2\pi}}\sqrt{\frac{l}{|x|}}\Sigma _{\pm}\int_0^\infty d\rho \frac{\sqrt{\varrho}}{\cosh(\rho\frac {x}{|x|}\frac{H}{l})}\\
\exp\big(\pm i t\sqrt{\frac {g\rho}{l}\tanh (\rho\frac{H}{l})}\big)\Sigma _{\pm}\big(
 e^{\mp i\pi/4} e^{\frac{{\pm}i\varrho|x|}{l}}\tilde V({\pm}\rho
\frac{x}{|x|})\big).
\end{gather*}
%Знаки $\pm$ в последнем интеграле выбираются независимым образом.
Let us consider the last integral. Its complete phases are:

$$\Phi_{\pm,\pm}/l=\pm(t\sqrt{gl\rho\tanh(\rho\frac{H}{l})}{\pm}\rho|x|)/l.$$

For $t>0,\rho >0$ the derivative
$\frac{\pa \Phi_{\pm, +}}{\pa \rho}$ is strictly positive, this implies
the absence of critical points for the functions $\Phi_{\pm,
+}$. It follows that these terms give, for $t>0$, a
contribution to the wave field which is asymptotically small with
respect to the other contributions and so it can be dropped.
Furthermore since $V$ is a real function then $\tilde V(\rho
\frac{x}{|x|})$ and $\tilde V(-\rho\frac{x}{|x|})$ are complex
conjugates so the last integral may be written in the form
\begin{gather*}
\eta\approx\frac {1}{\sqrt{2\pi}}\sqrt{\frac{l}{|x|}}\times\\\rm{Re}
\int_0^\infty d\rho \frac{\sqrt{\varrho}}{\cosh(\rho\frac {x}{|x|}\frac{H}{l})}\tilde V(\rho
\frac{x}{|x|})
e^{-i\pi/4} \exp\big(\frac {i}{l}(\varrho|x|-
t\rho\sqrt{g H}\sqrt{ \frac{l}{\rho H}\tanh (\frac{\rho H}{ l}) }\big).
\end{gather*}
Since the ratio $\frac{H}{l}$ is rather small, the source is
localized, and the function $\tilde V(\rho \frac{x}{|x|})$
decays rapidly as a function of  $\rho$, then the main
contribution to the last integral is coming from the small values
of $\rho$. Then we get that the functions $\frac{1}{\cosh(\rho\frac
{x}{|x|}\frac{H}{l})}$ and $t\rho\sqrt{g H}\sqrt{ \frac{l}{\rho
H}\tanh (\frac{\rho H}{ l})}$ can be expanded in Taylor series. If
we substitute the first function with $1$ we neglect a term of the
order of $O(\frac{ H}{ l})^2$. The second function can be
approximated by the first two non zero terms of its expansion
$t\rho\sqrt{g H}(\frac{1}{l}-\frac {1}{6}(\frac {\rho H}{l})^2) $
making an error of the order of $t\sqrt{g H}(\frac{H}{l})^4 $. It
is clear from the previous estimates that these terms are small
and so we obtain
\begin{gather*}
\eta\approx\frac {l}{\sqrt{2\pi}}\sqrt{\frac{l}{|x|}}\rm{Re}
\int_0^\infty d\rho {\sqrt{\varrho}}\tilde V(\rho \frac{x}{|x|})
e^{-i\pi/4} \exp\big(\frac {i}{l}(\varrho|x|- t\rho\sqrt{g
H}(1-\frac {\rho^2}{6}(\frac { H}{l})^2) \big)\big).
\end{gather*}
It will be explained below that the integral gets its larger
values in the neighborhood of the front, i.e. near the curve (
circle) $|x|=\sqrt{g H}t$. In this way the dispersion effects can
have an influence on the asymptotic values in the far wave field
under the condition that the coefficient of $\rho^3$ in the
exponent is larger or equal to one. Thus we obtain different
behaviors, putting $\sqrt{gH}t$ equal to $|x|$ in this
coefficient, according to different relations among $|x|,H,{l}$
(compare \cite{Voit}-\cite{Ivanov},\cite{berry}):

{\bf a)} For $|x|>>\frac{l^3}{H^2}$ the dispersion has an important influence in the neighborhood of the front,
and the asymptotic can be expressed by means of a function similar to the Airy function. In this case the behavior
of the function $V$ is not important for the definition of the profile of the front.

{\bf b)} For $|x|\sim\frac{l^3}{H^2}$ the weak dispersion and the function $\tilde V$ have equal influence on the formation of the wave profile;

{\bf c)} For $|x|<<\frac{l^3}{H^2}$ the dispersion is not important. If the term with $\rho^3$, is dropped from the phase of the integral
an error of the order of $|x|H^2/l^3$ is done.

Let us consider the example where $H=4 km$, $l=40 km$, thus
$l^3/H^2=4000 km$.  Thus a (weak) effect of the dispersion starts
at $4000 km$. If the size of the source increases twice this
distance increases 8 times and becomes 32000 km, a distance larger
than any ocean. Thus we will start analyzing the point {\bf c} (it
possible to neglect the effect of the dispersion).

\subsection{Asymptotic behavior of the wave field with very small dispersion in the case of constant depth}

{\bf NEW}

Thus, assuming that the inequality $|x|<<l^3/H^2$ is satisfied, we have
\begin{gather*}
\eta\approx\frac {1}{\sqrt{2\pi}}\sqrt{\frac{l}{|x|}}\rm{Re} \int_0^\infty d\rho {\sqrt{\varrho}}\,\widetilde V(\rho
\frac{x}{|x|})
e^{-i\pi/4} \exp\big(\frac {i}{l}(\varrho|x|-
t\rho\sqrt{g H}\big)=\\
\frac {l^{1/2}}{\sqrt{|x|}}\rm{Re}\big(e^{-i\pi /4}F(\frac{\Phi(x,t)}{l},\frac {x}{|x|})\big),\qquad \Phi(x,t)=|x|-t\sqrt{g H},
\end{gather*}
where
\begin{equation}F(z,\mathbf{n} )= \frac {1}{\sqrt { 2\pi}}\int_0^{\infty} e^{iz\rho}
{\sqrt {\rho}\widetilde V(\rho \mathbf{n})d\rho} %f(\rho,\mathbf{n}}\,\big),\,\,\, f(\rho,\mathbf{n})=\frac {1}{\sqrt { 2\pi}}\widetilde V(\rho \mathbf{n}),
\quad \mathrm{n}=\frac{x}{|x|}.\label{F}\end{equation}
%and $z= \frac{|x|-t\sqrt{g H}}{l}$.
Here $\mathbf{n}$ is a unit vector. It is natural to introduce its angle $\psi$ in such a way that $\psi=0$ corresponds to the axis $x_1$. Then \begin{equation}\label{n}
\mathbf{n}=\mathbf{n}(\psi)=\begin{pmatrix}\cos \psi \\ \sin \psi \\
\end{pmatrix}.
\end{equation} Hence the function $\widetilde V(\rho,\mathbf{n}(\psi))$ depends on $(\rho,\psi)$ and the function $ F(z,\mathbf{n}(\psi))$ depends on $(z,\psi)$. Not to complicate notation we use the same symbols
$\widetilde V$ and $F$ for them and sometimes write  $\widetilde V(\rho,\psi)$ and
$F(z,\psi)$ instead $\widetilde V(\rho,\mathbf{n}(\psi))$ and $F(z,\mathbf{n}(\psi))$ respectively.

 We note, that the function $F(z,\mathbf{n})$ decreases for $|z|\to \infty $ as an inverse power. Indeed, let us change variable in the
 last integral $\rho=\frac{y^2}{2}$; then
\begin{equation*}
F(z,\mathbf{n})=\frac {1}{\sqrt{2\pi}}\rm{Re}\int _0^{\infty}\, y^2
\, e^{iz(\frac {y^2}{2}-\frac{\pi}{4})}\,\widetilde
V(\frac{y^2}{2}\mathbf{n})dy.
\end{equation*}

\noindent Let us use the method of the stationary phase, we get,
because of the presence of the factor $y^2$ under the integral,
 $F(z,\omega)\sim \frac{1}{z^{3/2}}$, if $\widetilde V (0)\not=0$. Thus for \break $\big| |x|-\sqrt {gH}t\big |>>l$ and
  $|x|>>l$, we have that $\eta \sim \frac {l^3}{|x|^2}\widetilde V(0)$.
%{\bf IMPORTANT: there is disagreement between formulas below, please check them!!!!}
\begin{exam}
Let us give some example of the function $F(z,\omega)$. We choose for the function $V$, defining the source, the function
\begin{gather}\label{Sour}
V(y)=\bar V \cos{(a_1 Y_1+a_2 Y_2+\chi)}
e^{-b_1 Y_1^2 - b_2 Y_2^2},\,\, Y=\Theta(\theta)%^{*}
y,
\\\nonumber
\Theta(\theta)=\begin{pmatrix} \cos\theta &\sin
\theta\\-\sin\theta& \cos \theta
\end{pmatrix},
 %\label{4.1}
 \end{gather}
where $\bar V,a_1,a_2,  b_1, \  b_2>0,
\,\theta,\chi$ are parameters. In this case 
the function $F(z,\psi)$ can be expressed in terms of 
parabolic cylinder functions $D_{-3/2}$ or  
confluent hypergeometric functions ${}_1F_1$  
(see \cite{RyzhGrad}, 3.462, page. 351)
\begin{gather}\label{Sour1}
\widetilde{V}(\rho,\psi)=
\frac {\bar V\sqrt{\rho} }{2 \sqrt{b_1 b_2}}
e^{- \alpha - \beta \rho^2}\cosh(i \delta + \gamma
\rho),\,\,\\\nonumber F(z,\psi)=\frac{\bar V\sqrt{b_1 b_2}e^{-\delta}}{2\sqrt{2\pi}}
\rm{Re}\big(e^{-\frac{i\pi}{4}}\int_0^\infty\sqrt{\rho}
\big(e^{-\frac{\rho^2\beta}{2}+\gamma\rho
+i\rho z}e^{i \theta}+e^{-\frac{\rho^2\beta}{2}-\gamma\rho
+i\rho z}e^{-i \theta}\big)d\rho\big)\equiv\\\nonumber
=\frac{\bar V\sqrt{b_1 b_2}}{4 e^{\delta}\beta^{3/4}}\rm{Re}\big( \exp(\frac {(\gamma+iz)^2}{4\beta})D_{-3/2}(-\frac{\gamma+iz}
{\sqrt \beta})e^{i \theta}+
\exp(\frac {(-\gamma+iz)^2}{4\beta})D_{-3/2}(-\frac{-\gamma+iz}{\sqrt \beta})e^{-i \theta}\big)\equiv\\\nonumber
\bar V \sqrt{\frac{1}{32 \pi b_{1} b_{2} }}
Re[( Q_{+} + Q_{-})],
\end{gather}
\begin{gather}\nonumber%\label{Sour2}
Q_{\pm}(z, \psi)
=\frac{e^{- i \frac{\pi}{4}-\alpha\pm i \delta
}}{(5 \pm 3) \beta^{5/4}} \ \Bigl(\mp \ {\sqrt{\beta}} \
\Gamma(\frac{1}{4} \mp \frac{1}{2}) \ \
            _{1}F_{1}(\frac{3}{4}, \ \frac{1}{2}, \ \frac{w_{\pm}^{2}}{4 \beta})  +
            w_{\pm} \ \Gamma(\frac{1}{4}) \ \
            _{1}F_{1}(\frac{5}{4}, \ \frac{3}{2}, \ \frac{w_{\pm}^{2}}{4 \beta})
            \Bigr),\end{gather}
where $\Gamma$ is a gamma function,  and
\begin{gather*}\nonumber w_{\pm}= \ \pm \gamma + i z,\ \
\ \
\alpha=%\frac {1}{4 b_1 b_2}
(b_1 a_2^2 + b_2 al_1^2)/({4 b_1 b_2}), \\
\ \  \beta=%\frac {1}{4 b_1 b_2}
(b_1 {\sin}^{2}(\psi-\theta) +
b_2 {\cos}^{2}(\psi-\theta))/({4 b_1 b_2}),\\\nonumber
\gamma=%\frac {1}{2 b_1 b_2}
(b_1a_2 \sin(\psi-\theta) + b_2 a_1 \cos(\psi-\theta))/(2 b_1 b_2),
\end{gather*}

%e^{-(\rho^2}e^{+(b_1\alpha_1\cos\psi+b_2\alpha_2\cos\psi)\rho} e^{-}=Ae^{-\rho^2\frac{\beta}{2}+c\rho}e^{-D}

%Then \cite{RyzhGrad}, 3.462, page. 351\begin{gather*}F(z,\mathbf{n}(\psi))=\frac{A\sqrt{b_1 b_2}e^{-\delta}}{2\sqrt{2\pi}}\rm{Re}
%\big(e^{-\frac{i\pi}{4}}\int_0^\infty\sqrt{\rho}\big(e^{-\frac{\rho^2\beta}{2}+\gamma\rho
%+i\rho z}e^{i \theta}+e^{-\frac{\rho^2\beta}{2}-\gamma\rho
%+i\rho z}e^{-i \theta}\big)d\rho\big)=\\
%=\frac{A\sqrt{b_1 b_2}}{4 e^{\delta}\beta^{3/4}}\rm{Re}\big( \exp(\frac %{(\gamma+iz)^2}{4\beta})D_{-3/2}(-\frac{\gamma+iz}
%{\sqrt \beta})e^{i \theta}+%\\
%\exp(\frac {(-\gamma+iz)^2}{4\beta})D_{-3/2}(-\frac{-\gamma+iz}{\sqrt %\beta})e^{-i \theta}\big),
%\end{gather*}
%where $D_{-3/2}$ are the parabolic cylinder functions.

%Inserting this functions in the formula for the amplitude we get

%\begin{gather}
%\eta \approx \frac{A b_1 b_2 e^{-\delta}}{\sqrt{32 \pi |x|}}
%\op{Re} \Bigl\{ e^{-i\frac{\pi}{4}}\int_{0}^{\infty}e^{i Z \rho}
% \sqrt{\rho} e^{-\beta (\psi) \rho^2}
  %\cr (1 + i \sigma \rho^3)
% \Bigl[e^{\gamma (\psi) \rho+ i \chi
%} + e^{- \gamma (\psi) \rho-i \chi }
%\Bigr]\ d\rho \Bigr\}  \big|_{Z=(|x|-t\sqrt{g H})/l}\label{6.3}
%\end{gather}
%where\begin{gather*}\sigma=\frac{t}{6} \sqrt{g h_0} h_0^2. %\label{6.2}\end{gather*}

The figures Fig. 1, Fig. 2  present some types of 
profiles. 
%\begin{figure}
%\begin{center}
%\includegraphics{Fig2.eps}
%\end{center}
%\end{figure}
%for the case 
%when $V=\bar V e^{-b_1 y_1^2 - b_2 y_2^2}$, $b_1=...$, $b_1=...$ for the angles $\psi=......$.
%The figure Fig.2 present the  the different types of profiles for the case when $V=\bar V V(y)=\bar V \cos{(a_1 Y_1+a_2 Y_2+\chi)}
%e^{-b_1 Y_1^2 - b_2 Y_2^2}$, $b_1=...$, $b_1=...$, $a_1=....., a_2=....$
%for the same angles $\psi=......$.  We choose  the angle $\theta=0$, it does not play any role in the case of a constant bottom. We see that the wave profiles and amplitudes depend not only on the characteristic of the source but also on the angle ``cutting'' the the source: for some direction the amplitude could be very small, and the number of oscillation can be different.
%{\bf  TIMUR!!!! I THINK HERE WE NEED to put  pictures illustrated these ideas(FIG.1,FIG.2).}
\end{exam}

{\it Main conclusion}: the phase in the neighborhood of the front
defines completely a one parameter family of trajectories which
generate the front. Further we remark that, since the function $F$
decreases in a neighborhood of the front, we can expand in the
formula (\ref{F}) $|x|$ in a neighborhood of the front, keeping in
the expansion only the zero order term, and that we can substitute
the factor
  $\frac {1}{\sqrt{|x|}}$ ( the amplitude of the wave) with the term  $\frac {1}{\sqrt{\sqrt{gH}t}}$.
We want to find analogous formulae for the wave field in the case of negligible small dispersion and for variable bottom.

\section{ Asymptotic behavior of the wave field over nonuniform bottom for very small dispersion}
\subsection{The wave equation, rays and wave fronts }
In this section we start the analysis of the behavior of the
amplitude of the wave when the bottom is not constant.
We use here well known objects and their characteristics 
which one can find in  books connected with the semiclassical 
asymptotics and ray method, geometrical optics and wave fronts, 
Hamiltonian mechanics, catastrophe theory etc. We try to collect 
here all necessary objects and give  their description in 
elementary form. More complete form and  details  one can find in
\cite{MaslovAsymptMethods}-\cite{Kiselev}    
 It is clear that in practice we have studied the
solution of the wave equation in the previous section. In order to
be accurate we introduce in the calculations the characteristic
depth $H_0$ of the basin,  the small parameter
\begin{equation}\label{mu}
\mu=\frac l L
\end{equation}
expressing the relationship among the characteristic size of the
source and the characteristic size of the basin. We begin
introducing non dimensional variables in the equations. Then after
a suitable change of variables $x=x'/L$, $t=t'\sqrt{gH_0}/L$,
$H=H_0H'(x')$ our equations and initial data will take the form:
\begin{gather}
\frac{\partial ^2\eta}{\partial t^2}=g<\nabla,H(x)\nabla>\eta,\label{WEq}\\
\eta|_{t=0}=V(\frac x \mu), \quad\eta_{t}|_{t=0}=0.\label{WEq1}
\end{gather}
Our asymptotic
expansions will be done in term of this parameter under assumption $\mu\ll 1$ and that the domain in $\mathbb{R}^2_x$ and time interval $[0,t]\in\mathbb{R}_t$  while asymptotic expansion are working independent of $\mu$.{\it To come back to original variables it is enough to change in final asymptotic formulas $\mu$ by $l$.}

We assume that the source of the perturbation is localized in
$x=0$. It is easy to see that finding the field far from the
source,$|x|\gg l$, is similar to find the asymptotic values for
$\mu\to 0$ in the problem (\ref{WEq}). The problem now is to study
the wave equation with variable coefficient. The asymptotic values
of the wave amplitude $\eta$ can be expressed by means of the wave
front formed by rays.  It is a known fact that instead of the
straight rays one has to introduce curved rays and characteristics
which are  the one dimensional family of trajectories
$P(\psi,t),X(\psi,t)$ of an appropriate Hamiltonian system . The
ends of the rays again form the wavefront, but now it can be
different from the circle, a more complicated closed curve
probably with cusps  and self intersection points. In the
considered situation these rays and  characteristics are
determined in the following way.

We introduce the function $C(x)=\sqrt{g H(x)}$ and, as before, let
$\mathbf{n}$  be the unit vector \eqref{n}
directed as the external normal to the unit circle. Then  the
Hamilton system is:
\begin{gather}
\dot{x}=\frac{p}{|p|}C(x),\quad
\dot{p}=-|p|\frac{\partial C} {\partial x},\qquad%\label{Ham2}\\
x|_{t=0}=0, \quad p|_{t=0}=\bf {n}(\psi),\label{Ham3}
\end{gather}
i.e. the family of trajectories $P(t,\psi),X(t,\psi)$ going out
from the point $x=0$  with unit impulse $p=\bf {n}(\psi)$. Let us
indicate $C(0)=C_0$. The Hamiltonian  corresponding to
(\ref{Ham3}) is $\mathcal{H}=C(X)|p|$. From the conservation of
the Hamiltonian on the trajectories we have the important equation
\begin{equation}|P|C(X)=C_0.\label{Ham4}\end{equation}
The projections $x=X(t,\psi)$ of trajectories on the plane $\mathbb{R}^2_x$
are called the rays. Recall that the {\it front} in the plane $\mathbb{R}^2_x$ at the
time $t>0$ is the curve $\gamma_t
=\{x\in\mathbb{R}^2|x=X(\psi,t)\}$, %\cite{Arnold,Maslov}. 
The  points on this curve are parameterized  by
the angle $\psi\in (0,2\pi]$.
  If in each point $x$ of the front
$\gamma_t\quad$ $\frac{\partial X}{\partial \psi}\not =0$, then
the front is a smooth curve. The points where $\frac{\partial
X}{\partial \psi}=0$ are named {\it focal} (backtracking or 
turning points), in these points the
front looses its smoothness. In the situation in which the focal
points appear, (they are very interesting from the point of view
of tsunami), it is reasonable to introduce the concept of the
front in the phase space $\mathbb{R}^4_{p,x}$ at the moment $t>0$,
i.e. the curve $\Gamma_t=\{p=P(\psi,t), x=X(\psi,t), \psi\in [0,2\pi]\}$.
We note that at least one of the component of the vector $P_\psi, X_\Psi$
is different from zero, see Lemma 3; 
and also  the  rays $x=X(t,\psi)$ are 
orthogonal to the front $\gamma_t$: 
$\langle \dot X,X_\psi\rangle=0$ see Lemma 3.

\subsection{The  wave field  before critical times.}
It is not difficult to check that a (possibly sufficiently small)
$t_1$ exists  such that, for any $t$, $t_1\geq t
>\delta>0$, there are no focal points in $\gamma_{t}$. The first instant
of time $t_{cr}$, in which focal points are formed is called {\it
critical}. Let us first write the solution before critical times,
larger than $\delta$, when the front is already defined. In this
case the asymptotic solution is defined in the following way. We
define a neighborhood of the front for sufficiently small ( but
independent of $h$) coordinates $\psi,y$, where $|y|$ is the
distance among the point $x$ belonging to a neighborhood of the
front and the front. For this aim we will take $y\geq 0$ for the
external subset of the front and  $y\leq 0$ and for the internal
subset of the front. Then a point $x$ of the neighborhood of the
front is characterized by two coordinates: $\psi(t,x)$ and
$y(t,x)$, where $\psi(t,x)$ is defined by the condition that the
vector $y=x-X(\psi,t)$ is orthogonal to the vector tangent to the
front in the point $X(\psi,t)$. Thus we have the condition
$\langle y,X_\psi(\psi,t)\rangle=0$.  Let us find the phase
$$S(t,x)=\langle
P(\psi(t,x),t),x-X(\psi(t,x),t)\rangle=
\frac{C(0)}{C(X(\psi(t,x),t))}y=
\sqrt{\frac{H(0)}{H(X(\psi(t,x),t))}}y$$

The second equality is a consequence of the equation (\ref{Ham4}).

%NEW {\bf Comments to the changes: I think that the function $F$ should be substituted by the function $V$, which models the initial perturbation, but then the arguments of the function should be written in some other way since $V$ is a function of the vector $x$. In any case it is not clear the connection between $F$ and $V$. The phase $\Phi$ should be substituted by the phase $S(x,t)$. I changed also some English expressions and words}

Now we state the first important theorem of this paper connecting
the wave amplitude with the initial perturbation $V(x)$ and the
profile of the bottom and the integration over the
characteristics.

%{\bf Sergey's  Comments: I agree with you, but we need to check all formulas more carefully. There is also a question about parameter $h$. In "Doklady"  we make all consideration in dimension variables and $h$ appear as $\rho/l$.  Thus the dimensionless parameter $\mu=l/L$, and in the final formulas (in Can.Operator)  we need to change $h$ by $\mu$ or vise verse.}

%{\bf New}
\begin{theorem} For  $t_{cr}>t>\delta>0$ in some neighborhood of
the front $\gamma_t$, not depending on $\mu$, $ \eta $, the
asymptotic elevation of the free surface, has the form:
 \begin{gather}\eta= \frac{\sqrt{\mu}}{\sqrt{|X_{\psi}(\psi,t)|}}\sqrt[4]
{\frac{H(0)}{H(X(\psi,t))}}\op{Re}\big[e^{-\frac{i\pi}{4}}
F\big(\frac{S(t,x)}{\mu},
{\mathbf{n}(\psi)}\big)\big]\big|_{\psi=\psi(t,x)}+O(\mu^{3/2}). \label{WF}
\end{gather}
Outside this region $\eta=O(\mu^{3/2})$. The function $F(z,\mathbf{n})$ is defined in \eqref{F}.
\end{theorem}

In this way till the critical time the asymptotic elevation of
the free surface is completely defined by means of the trajectory,
which forms the front of the wave, and of the function $V$,
corresponding to the source of the perturbation. Despite of the
simple and natural form of the asymptotic of $\eta$, the proof of
the formula (\ref{WF}) is not trivial at all; the main step is the
computation of the function $V$, more exactly  the proof of the
fact that the formula is the same as in the case of constant
bottom, if the right choice of the rays is made. We will give
below a constructive approach of the proof of this formula, in the
meantime we now show some elementary consequence of the equation
(\ref{WF}). Since the phase $S(x,t)$ is equal to zero on the front
and $S(x,t)/\mu$ gets large going out from the front, then $\eta$,
as one could expect, decreases enough quickly and the maximum of
$|\eta|$ is attained in a neighborhood of the front. As a
consequence, $\eta$ can have some oscillations depending on the
form of the source. The second factor in (\ref{WF}) is the two
dimensional analogue of the Green  law, well known in the theory
of water waves in the channels: the amplitude $\eta$ increases when the depth
decreases as the inverse of the fourth root of the depth
$1/\sqrt{C(x)}=1/\sqrt[4]{H(x)}$; the factor $1/\sqrt{|X_\psi|}$
is connected to the divergence of the rays, in other words if a
smaller number of rays goes through a neighborhood of the point
$X(\psi,t)$, the smaller will be the amplitude of the wave field.
The factor $\frac{C_0}{C(X(\psi(t,x),t))}$ appearing in the
formula of the phase expresses the phenomena, also well known, of
the ``contraction'' of the wave profile as the depth decreases and
the increase of its amplitude. In fact the amplitude increases
because of the factor in front of the function $V$ but also the
phase $S(x,t)$ increases and this makes the wave profile narrower.
This result explains the well know fact that the wave length of
the tsunami decreases when the wave approaches the coast and that
its amplitude increases. The same profile (i.e. a section of
$\eta(x,t)$ for fixed $t$ and $\psi$) can depend on the way the
trajectory ( ray) intersects the initial perturbation of the
bottom at $t=0$. It is just this fact to give the dependence of
the diagram of the directions on two factors: the shape of the
source and the angle of its intersection with the ray passing
through a given point of the front. For this reason, depending on
the form of the bottom, two rays going out with two very different
angles, can arrive near a point of the front and contribute to the
profile with very different amplitudes. These effects can be well
seen in the figure Fig.3
\begin{rem}
The main argument usually used for deriving the the analytical and
and asymptotic formula of the solution of \eqref{WEq} is that,
after the front is formed, the problem is essentially one
dimensional in space till the appearance of focal points and its
dynamic is described by the one dimensional wave equation with non
uniform velocity. But it is possible to construct different
particular wave solution, localized in different neighborhoods of
the same front but having completely different profiles. The
question is the right choice of the function describing this
profile, i.e. a function connected already with the construction
of the solution of bi-dimensional problems, containing
mathematical difficulties such as the presence in the solutions of
the wave equation ( with variable coefficients) of effects of
intersection of the characteristics (which in quantum mechanics are also called terms). This happens for
very small wave vectors, i.e. only for very long wave lengths. A
general approach allowing to treat difficulties of this type were
developed in \cite{MaslovOperMethods} and, in particular for the Cauchy problem with
localized initial conditions for the wave equation and hyperbolic
systems, in the papers \cite{DobrZhShaf,DobrMZhSh}. These approach is basically founded on
the following argument. The solution has two contributions, the
first, corresponding to "very long" waves, can be found only
directly by numerical methods. It has a very small amplitude and,
in the case of the problem of the tsunami does not have a very
great importance. The second has a wave length long compared with
the depth of the basin of the wave but sufficiently small compared
with oceanic scale, in such a way that it is possible to apply
effective asymptotic methods as the expansion in rays. These
arguments for the given problem \eqref{WEq} were given accurately
in the papers \cite{DobrZhShaf,DobrMZhSh}, but the final formulas, based on the asymptotic
for the system with constant coefficients \cite{MasFed1}, are not
very efficient from the point of view of practical applications.
The basic of the derivation of the formula \eqref{WF} of
\cite{MasFed1} consists of two steps: 1) the construction of the
asymptotic expansion in the smoothness of the fundamental
solutions of the problem \eqref{WEq} ( parametrix) and 2) the
evaluation of the asymptotic of the convolution of this solution
with the initial function $\eta_0$. Basically our main 
observation (
missing in the papers \cite{MasFed1}) is that this last asymptotic
may represented in a simple and useful way \eqref{5.3}, from which
the representation \eqref{WF}, as well as representations \eqref{5.2}, \eqref{etafoc} working after appearance of critical points follow
immediately.
\end{rem}

%{}\vspace{1cm}

\subsection{The structure and metamorphosis of wave
profiles.}

\subsubsection{The  Maslov index and metamorphosis of
the wave profile.$\label{fronts}$}

%{\bf It seems the case to join together all the different parts of the paper where the Maslov is treated, it is too much scattered. So it is the case to join the subsections $2.5.1$, together with the example and without the theorem, section $2.6$, $2.7.1, 2.7.2, 2.7.3 $, together with the examples.Perhaps also the subsection about the Maslov index of singular maps must be included in this big section. So there must be a section which includes the various subsections about the Maslov index, one about the scattering amplitudes, another one about the final formula for the amplitude.} {\bf NNEW: Sergey's Comment  I have repaired globally the last sections starting from Sec.5  }

 For $t>t_{\rm{cr}}$ when the
focal points appear, as it is well known  in the wave theory, the
front can have the ``angles'' and sometimes the front lines can
have self intersection points. The ends of the arcs corresponding
to these angles are the {\it focal points} (or backtracking or turning points). For  $t>t_{\rm{cr}}$
the front divides in some arcs $\gamma_t^j$, indexed by the number
$j$, separated by focal points. The internal points of these arcs
are the ends of the trajectories $P(\psi,t),X(\psi,t)$ with the
same topological structure. Namely these equivalent trajectories
cross the same numbers of focal points at times $t^F$ before $t$,
$t^F<t$. They are characterized, from the topological point of
view, by the {\it Maslov index}, {\it an integer number}
$m(\psi,t)$ {\it depending on} $\psi,t$. The Maslov index $m$ can
be defined on the {\it regular} points of the front in different
ways, we give below a more practical definition of this important
concept by a simple definition of its increments in the problem
under examination. The index $m$ is related to the sign of the
Jacobian $J=\frac{\pa X}{\pa(t,\psi)}\equiv (\dot X,X_{\psi})$.
%{\bf Sergey: this ordering in the definition of J is true!!!!}
The function $J$ is equal to zero in the focal points and only in
these points. Thus moving along the front $\gamma_t$ or along the
trajectory $(P,X)$ after crossing the focal point, the Jacobian can
change its sign. Actually the Maslov index prescribes a receipt
for assigning the correct sign to the square root of $J$ and it
can be defined in a way independent from the trajectories. But if
we move along a trajectory there is, in this problem, the nice and
useful fact that the Maslov {\it index coincides with the {\bf
simpler} Morse index}. So, considering the trajectories arriving
to $\gamma_t^j$, we have that {\it the Morse index $m(\psi,t)$ of
the point $x=X(\psi,t)\in \mathbb{R}^2_x$ is equal to the number
of focal points on the trajectory
$p=P(\psi,\tau),X(\psi,\tau),\tau\in(0,t)$ arriving to
$x=X(\psi,t)$.} Note also that, as the time $t$ changes, the ends
of the arcs $\gamma_t^j$ produce the entire set of focal points.
It is also a well known fact that these sets constitute the
(space-time) {\it caustics} which are the singularities of the
projections of some Lagrangian manifold (we denote it $M^2$) from
the phase space $\mathbb{R}^4_{p,x}$ to the plane
(configuration space) $\mathbb{R}^2_x$.

\begin{exam} $\label{bank1}$

Let us illustrate the concepts explained above by the example
\cite{} about the waves on an axially symmetrical bank described
by the the depth function
\begin{gather}
H=H(\rho),\,\rho=\sqrt{x_1^2+x_2^2}.\label{4.1}
\end{gather}

In this case there exists an additional integral
\begin{gather} p_\varphi=x_1 p_2- x_2 p_1 \label{4.2}\end{gather}
and the Hamiltonian system (\ref{4.1}) is completely integrable.

We assume that the source is located in a neighborhood of the
point $x_1=0, x_2=-\rho_0$.  After the appearance of the focal
points one has the following picture (see fig.4)%{\bf HERE it is good to put figure(see fig.4)}. 
For each fixed
time $t$ the front $\gamma_t$ is separated into two arcs: the
first, a long one, is $\gamma_t^1$ with self-intersection, and the
second, a short one, is $\gamma_t^2$, located between the angles
on the fronts. The union of the ends of the arc $\gamma_t^2$ for
different times $t$ gives a {\it caustic}. The arc $\gamma_t^1$
consists of the ends of trajectories (rays) without focal points
on them (except $t=0$). Thus  the Jacobian $J(\psi,
t)=\rm{det}\big (\dot X,X_\psi)(\psi,\tau)>0$ for fixed $\psi$ and
for each $\tau\in (0,t]$; hence the Morse index
$m(x\in\gamma_t^1)=0$. On the contrary the arc $\gamma_t^2$
consists of the final points of the trajectories (rays) which
cross one focal point at some time $t=t_{F}(\psi),
0<t_{F}(\psi)<t$ when they touch the caustic. In this case
before $t_{F}(\psi)$ $J>0$, $J(\psi, t_{F}(\psi))=0$, and $J<0$
for $t>t_{F}(\psi)$. Hence $m(x\in\gamma_t^2)=1$.
\end{exam}

%{\bf Questions: 1) why $J$ is negative for $t>t^F(psi)$? From what said just before this example about Maslov index coinciding with the Morse index $m=1$ just because a focal point has been crossed by the trajectory and so the value of the index seems not to depend on the sign of $J$! The fact is that all these questions are made clear with the Lemma 7 of the subsection 2.7.3, so here there should be a ref to that lemma}
%{\bf Sergey's Comment: the Jacobian changes its sign for sure,  and it was positive before the focal point. Thus it is negative after the focal point.}

Now let us  fix the time $t$ and  move along the front $\gamma_t$.
Then after the passage through the focal points the phase $-\pi/4$
in formula \eqref{WF} increases by a quantity $-\pi/4\pm\pi/2$,
where $\pm1$ is the jump of the Maslov index. Finally after
passing through several focal points instead of the factor
$e^{-\frac{i\pi}{4}}$ one has the factor
$e^{-\frac{i\pi}{4}-\frac{i\pi m(\psi,t)}{2}}$. The number $m$ is
defined $\rm{mod} \,4$. The appearance of this new factor {\it
produces crucial changes of the form of the wave profile} in the
formula \eqref{5.2} i.e. in the function
$\rm{Re}(e^{-\frac{i\pi}{4}-\frac{i\pi m}{2}}F)$. This fact is
analogous to the well known metamorphosis of the discontinuity in
the theory of hyperbolic systems, %\cite{} 
and
the formula \eqref{5.2} describes explicitly the appearance of the
same fact in the case of localized initial perturbations.

\begin{exam} %{\bf SERIOZHA I TIMUR: SJUDA ESHE KARTINKA}
 Let us
give the examples of the transformation (metamorphosis) of the
wave profile depending on the index $m$ and on the source of
gaussian type. We have the following pictures for $m=0,1,2,3,
\quad \op{mod}4$ (see fig ????). Thus in the cases $m=2,3$ we have
the ``overturned'' profiles with respect to the cases $m=0,1$. In
the considered example of formula \eqref{5.2} one gets the profile
(Fig.  ) for the long arc $\gamma_t^1$ and the profile (Fig.2) for
the short arc $\gamma_t^2$.
\end{exam}
Let us present the formula for the wave amplitude in a
neighborhood of the front but outside of some neighborhood of the
focal points. As we have just seen in the previous example, points
of self-intersection can appear for $t>t_{\rm{cr}}$. The amplitude
of the wave in a point $x$ belonging to a neighborhood of these
points now is the sum of the contributions coming from different
$\psi_j(x,t)$, $y_j(x,t)$, and $S_j(x,t)$ with index $j$, and with the  Maslov index  $m(\psi_j(x,t),t)$.
%{\bf NEW}
\begin{theorem}\label{1}
In a neighborhood of the front but outside of some neighborhood of
the focal points the  wave field is the sum of the fields
\begin{gather}\eta%(x,t)
 =\sum_j \{\frac{\sqrt{\mu}}{\sqrt{|X_{\psi}(\psi,t)|}}
\root 4 \of {\frac{H(0)}{H(X(\psi,t))}}\op{Re}\big[e^{-\frac{i\pi}{4}-\frac{i\pi m}{2}}
F(\frac{S_j(x,t)}{\mu},\mathbf{n}(\psi))\big]\}\big|_{\psi=\psi_j(x,t)}+ O(\mu^{3/2}).
\label{5.2}\end{gather} Outside this neighborhood of the front
$\gamma_t\,\,$ $\eta(x,t)=O(\mu^{3/2})$. Again the function $F(z,\mathbf{n}(\psi))$ is determined in \eqref{F}.
\end{theorem}
%{\bf ENDNEW} {\bf SERIOZHA I TIMUR, I THINK IT WILL BE GOOD TO GIVE A PICTURE FOR THE WAVEFIELD WITH THE AXILLY SIMMETRIC BANK: THE SUM OF THE PROFILES NEAR THE SELF-INTERSECTION POINTS}

Let us emphasize that the number $m$ has a pure topological and
geometrical character and can be calculated without any relation
with the asymptotic formulas for the wave field. From the theorem
\ref{1} it follows that, in order to construct the wave field  at
some time $t$ and in a point $x$, one has to know only the initial
values $\eta|_{t=0}$ and $\eta_t|_{t=0}$ and has not to know the
wave field $\eta$ for all previous time between $0$ and $t$. The
trajectories and the Maslov (Morse) index take into account all
metamorphosis of the wave field during the evolution from zero
time until time $t$.
\begin{rem}
Finally let us note that it  is possible to
define the Maslov index in the singular (focal) points also (see
\cite{MaslovTeorVoz}), actually this definition is associated with
a chain of covering maps of the front and gives the correct wave
field asymptotic in a neighborhood of the focal point. It is
useful to distinguish these two types of  Maslov indices; we also
meet in our calculations the second one, denoted by $\mathbf{m}$,
but after the discussion of the index $m$ (see subsection
(\ref{maslovmorse})).
\end{rem}

%{\bf NEWNEW I have rewritten the next SECTION almost completely}
\subsection{  Wave field asymptotic  in a neighborhood of focal point}
\subsubsection{Completely nongenerated focal points and coordinate system}
Now we consider the situation when for some $t$ the point $(P^F,X^F)=(P(\psi^F(t),t),X(\psi^F(t),t))$
corresponding to the angle $\psi^F(t)$ is a {\it focal} one. In this point $X_\psi$=0 and one has to use another
asymptotic representation for the solution. %{\bf NNEW 
Roughly speaking the neighborhood of the point $X(\psi^F(t),t)$ on the plane $\mathbb{R}^2_x$ can include several arcs of $\gamma_t$ with the angles $\psi$ far from $\psi^F(t)$. This means that one has to take into account contribution of all of these arks in  the final formulas for $\eta$ in the neighborhood of the point $x=X(\psi^F(t),t)$. The influence of nonsingular points are defined by formula
\eqref{5.2} and the influence of the points from the neighborhood of the focal points are described by formulas \eqref{etafoc} given below. Thus it is necessary to innumerate the focal points with the closed projection and write $P(\psi^F_j(t),t),X(\psi^F_j(t),t)$. These points have the closed position  $X^F=X(\psi^F_j(t),t)$, but different momentum $P^F=P(\psi^F_j(t),t)$. To simplify the notation we discuss here the influence into $\eta$  only one focal point omitting subindex $j$ but keeping $P^F$.

We present the corresponding formula  under
the assumption that some derivative.% ENDNNEW }
\begin{equation}\label{der}
X^{(n)F}_\psi=\frac{\pa^n X}{\pa \psi^n}(\psi^F(t),t)=\frac{\pa^n X}{\pa \psi^n}(\psi^F(t),t)\neq 0,
\end{equation}
and the derivatives $X^{(k)F}_\psi=0$ for $1\leq k<n$. It means that this
focal point is not completely degenerate.
For future it is convenient to introduce the ``mixed'' Jacobian
\begin{equation}\label{Jtilde}
\tilde J=\det (\dot X, P_\psi)(\psi,t)=\frac{C^2(X)\det (P, P_\psi)}{C_0}(\psi,t)
\end{equation}
and some characteristics of the focal point $(P^F,X^F)$:
\begin{gather}\nonumber
C_F=C(X^F),\,\,\, \dot X^F=\dot X (\psi^F(t),t)=\frac{P^F C_F^2}{C_0},\,P_\psi^F=P_\psi(\psi^F(t),t),\\
\tilde J_F=\det (\dot X^F, P_\psi^F)=\frac{C^2_F\det (P, P_\psi)}{C_0},\,\,J^{(n)}_F=\det (\dot X^F, X^{(n)F}_\psi).\label{foc}
\end{gather}

Again  the
topological characteristic appears, i.e. the Maslov index of this focal point
or its neighborhood (it is the same), but now it depends on the
choice of the coordinates in the neighborhood of  $(P^F,X^F)$. It
is natural to choose the new coordinates $(x'_1,x'_2)$ associated with the nonzero
vector $\dot X^F=\dot X(\psi^F(t),t)$; namely we assume that the
direction of the new vertical axis $x'_2$ coincides with the vector
$\dot X^F$. We put $\mathbf{k}_2={}^t(k_{21},k_{22})=\dot
X^F/|\dot X^F|$=$\dot
X^F/C_F={P^F C_F}/{C_0}$,
$\mathbf{k}_1={}^t(k_{11},k_{12})=(k_{22},-k_{21})$ and introduce
the  new coordinates $p',x'$ in the neighborhood of $(P^F,X^F)$ in the phase space $\mathbb{R}^4_{p,x}$ by
formulas:
\begin{gather}\nonumber
x'_1=\langle \mathbf{k}_1,x-X^F\rangle=-\frac{\det (\dot X^F,x-X^F)}{C_F}=-\frac{C_F}{C_0}{\det (P^F,x-X^F)},\\\nonumber
 x'_2=\langle
\mathbf{k}_2,x-X^F\rangle=\frac{\langle
\dot X^F,x-X^F\rangle}{C_F}=\frac{C_F}{C_0}{\langle
P^F,x-X^F\rangle},\\ p'_1=\langle
\mathbf{k}_1,p%-P^F
\rangle,\,\, p'_2=\langle
\mathbf{k}_2,p%-P^F
\rangle.\label{change}
\end{gather}
%{\bf NEW}
Easy to see that \begin{equation}\label{deter}
\det \begin{pmatrix}
  \dot P'_1 & P'_{1\psi } \\
  \dot X'_2 & X' _{2\psi }
\end{pmatrix}=\tilde J_F.
\end{equation}
\subsubsection{Maslov index of the focal point.}
%{\bf ENDNEW}
As the determinant $\tilde J\neq 0$ in the focal point $(P^F,X^F)$,
hence the same inequality takes place in some its neighborhood,
thus $\tilde J$ has a constant sign. On the contrary the Jacobian
$J$ changes  sign in this neighborhood. \emph{We define the Maslov
index $\mathbf{m}(P^F,X^F)$ of the non (completely)
degenerate focal point $(P^F,X^F)=(P,X)(\psi^F(t),t)$ as the  index $m(\tilde P,\tilde X)(\psi,t)$ of a regular point $(\tilde P,\tilde
X)=(P,X)(\tilde\psi,\tilde t)$ in the neighborhood of $(P^F,X^F)$
such that the signs of the determinants $J(\tilde\psi,\tilde t)$
and $\tilde J(\tilde\psi,\tilde t)$ coincide.} For instance one
can choose $\tilde \psi=\psi^F(t),\tilde t=t\pm\delta $, where
delta is small enough. This means that we compare the sign of $J$
with the sign of $\tilde J$ on the trajectory $(P,X)$  crossing
the curve $\Gamma_t$ in the focal point $(P^F,X^F)$ before and
after this crossing.
%{\bf this definition is not clear: if the sign of $J$ changes and the one of $ \tilde J$ remains the same how can one choose the regular points such that the signs are the same? There are always such regular points if $J$ can be positive and negative and so then $m$ is always equal to $1$! There also something not clear with the previous formulation of this example: why $J<0$ after crossing the focal point as it is written there? }
%{\bf Sergey's Comment. In the singular map there exists nonsingular points with two different signs, because $J$ changes it sign along the trajectory. The sign of $\tilde J$ is constant. Hence there exists the regular point in this map, where $J\tilde J>0$.}
\begin{exam}$\label{bank2}$
Let us illustrate this definition for the focal
points of the example with  the axial symmetric bank
\eqref{4.1}(see Fig...). We saw that before crossing the focal
points $J>0$, and after crossing $J<0$ and the index of the points
on this trajectory is $1$.
 Let us find the sign of $\tilde J_F$ in the focal
point. To this end we differentiate the integrals of motion
\eqref{4.2} of the Hamiltonian system with respect to the angle
$\psi$ and put $\psi= \psi^{F}(t)$.  We find
$$\langle P^F,P_\psi^F \rangle =0,\qquad X_2^F P_{1\psi}^F-X_1^F P_{2\psi}^F=\rho_0\sin \psi^{F}(t)
\qquad\text{for}\quad \psi= \psi^{F}(t).$$
The solution of this equation is
 $$\begin{pmatrix}P_{1\psi}^F\\P_{2\psi}^F\end{pmatrix}=
\frac{\rho_{0}\sin \psi^F(t)}{\langle P^F,X^F\rangle}(\psi^{F}(t),
t)\begin{pmatrix}P_{2}^F\\-P_{1}^F\end{pmatrix}$$ and
$\frac{|P^F|}{C_F}\rm{det}\big(\dot
X^F,P_{\psi}^F\big)=\rm{det}\big(P^F,P_{\psi}^F\big)=-\frac{\rho_0
P^2}{\langle P^F,X^F\rangle}\sin \psi^{F}(t).$ Obviously the angle
$\psi^F(t)$ belongs to the interval $(0,\pi)$ and ${\langle
P^F,X^F\rangle}>0$, hence $\tilde J_F<0$ and
$\mathbf{m}(P^F,X^F)=1$.

\end{exam}

\subsubsection {The model functions and the wave profile in neighborhood of the focal point.}
Now we present the formulas for the wave field in the
neighborhood of a focal point $x=X^F$. Let us put $\sigma=\sign(\tilde J_F J^{(n)}_F)$ and introduce the function (or more precisely  the linear operator acting to the source function  $V(y_1,y_2)$)
\begin{gather}\nonumber
g_n^\sigma(z_1,z_2,\psi)=
\int_{-\infty}^{\infty}d\,\xi \int_{0}^{\infty}\rho d \rho\tilde V(\rho \mathbf{n}(\psi))
\exp\{{i\rho}\big(z_2-\xi z_1-\sigma \frac{\xi^{n+1}}{(n+1)!}\big)\}=\\\int_{-\infty}^{\infty}d\,\xi \int_{0}^{\infty}\sqrt{\rho} d \rho\tilde f(\rho \mathbf{n}(\psi))
\exp\{{i\rho}\big(z_2-\xi z_1-\sigma \frac{\xi^{n+1}}{(n+1)!}\big)\}.
\label{gn}\end{gather}
We put
$$z_1^F=\frac{x'_1}{\mu^{\frac{n}{n+1}}}\frac{\tilde J_F }{|\tilde J_F J^{(n)}_F|^{\frac{1}{n+1}} C_F^{\frac{n}{n-1}}}\equiv-\frac{\det(P^F,x-X^F)}{C_0 C_F^{\frac{1}{n-1}}\mu^{\frac{n}{n+1}}}\frac{\tilde J_F }{|\tilde J_F J^{(n)}_F|^{\frac{1}{n+1}} },
\quad z_2^F=
\frac{x'_2}{\mu}\frac{C_0}{C_F}\equiv\frac{\langle P^F,x-X^F\rangle}{\mu}.$$
\begin{theorem} Each focal point $(P^F,X^F)$ on the front $\gamma_t$ gives in its neighborhood the following  contribution
\begin{equation}\label{etafoc}
\eta^F=\mu^{\frac{1}{n+1}}\{\frac{\sqrt{C_0
|\tilde J_F|^{\frac{n-1}{n+1}}}}{|J^{(n)}_F|^{\frac{1}{n+1}}C_F}
\rm{Re}[e^{-i\frac{\pi}{2}\mathbf{m}(P^F,X^F)}g_n^\sigma(z_1^F,z_2^F,\psi^F)]+
O(\mu)\}
\end{equation}
into the asymptotic of solution $\eta$. If the several arcs of $\gamma_t$ belong to  the neighborhood of the point $x$, then one need to summarize all corresponding functions \eqref{etafoc} and \eqref{5.2}.
\end{theorem}
%ENDNEW
\section{The geometric base of asymptotics: Lagrangian manifolds, the Maslov and Morse indices.}
The aim of the next section is to prove Theorems 1-3.  But let us
first recall the geometrical objects and the important properties
of the Hamiltonian system (\ref{Ham3}), giving an uniform
asymptotic solution to problem (\ref{Ham3}) including the behavior
in a neighborhood of focal points, initial moment of time,
calculation of the Maslov and Morse indices etc. The majority of
these constructions and properties are well known, we present them
in the most simple form and collect them in our paper for giving a
self-contained treatment. An exhaustive description of the wave
fronts and the focal points, their connection with the ray method
and the semiclassical asymptotic, can can be found for instance in
mentioned above monographs and papers. 
%\cite{Arnold,BabichBuldMolot, Maslov,MaslovTerVozm,
MaslovFed,Berry}. There exist different equivalent definitions of
the Maslov index;  
one of the aims of the next subsection is to recall
the definition from \cite{MaslovOpMethods,Arnold1,DobrZh} which, in our
opinion, is more suitable for concrete calculations.

\subsection{\bf  Lagrangian manifolds (``bands'') and their properties.}
As we have already said, taking into account the fact that after
the appearance of the focal points the front line can intersect
itself, it is convenient to add to  the point $x=X(\psi, t)$ the
corresponding momentum component $p=P(\psi, t)$, and consider the
point $\mathbf{r}=\mathbf{r}(\psi, t)=(P(\psi, t),X(\psi, t))$ in
the $4$ dimensional phase space $\mathbb{R}^4_{p,x}$. Each point
$\mathbf{r}(\psi, t)$ is completely defined by its coordinates,
which are the angle   $\psi$ (defined $\mod 2 \pi$) and the ``propper
time'' $t$.

Fixing the angle $\psi$ we obtain the trajectories
(bi-characteristics) of Hamiltonian system (\ref{Ham3}) in the
phase space $\mathbb{R}^4_{p,x}$, and, fixing the time $t$, we
obtain the front $\Gamma_t$ in the in the phase space
$\mathbb{R}^4_{p,x}$. The projections of the trajectories from
$\mathbb{R}^4_{p,x}$ to the configuration space (plane)
$\mathbb{R}^2_{x}$ are the {\it rays}. The projection of the curve
$\Gamma_t$ from $\mathbb{R}^4_{p,x}$ to the configuration space
(plane) $\mathbb{R}^2_{x}$ are the fronts $\gamma_t$. Different
points $\mathbf{r}(\psi_j, t)$ on $\Gamma_t$ can have the same
projection $x=X(\psi_j, t)$ on $\gamma_t$, but now we distinguish
them by different angles $\psi_j$.

Let us fix some small but independent of $\mu$ number $\delta$. According to \cite{Maslov,MaslovFed} changing both the angle
$\psi$ and the time $\tau\in (t-\delta,t+\delta)$ on the cylinder $\mathbb{S}\times
(t-\delta,t+\delta)$ we obtain the 2-D Lagrangian manifold (with the boundary) $M^2_{t}=\{p=
P(\psi, \tau), x=X(\psi, \tau)| \psi\in \mathbb{S},\tau\in (t-\delta,t+\delta)\}$;
the angle $\psi$ from the unit circle $\mathbb{S}$  and the time
$t$ from $(t-\delta,t+\delta)\in \mathbb{R}$ are the coordinates on $M^2_t$, sometimes we
shall use the notation $\alpha=\tau-t$ instead of the time  $t$. Actually the manifold $M^2_t$ has a structure of a  cylindrical ``band'' (or closed strip) with the width $2\delta$, thus we call it {\it Lagrangian band}; of course it depends on $\delta$, we omit this dependence to simplify the notation.
The family of Lagrangian bands  $M^2_t$ is {\it invariant} with respect to the phase flow $g^t_{\mathcal{H}}$ generated by the system
(\ref{Ham3}). This means that the point $\mathbf{r}(\psi_j,
\tau)$ from $ M^2_{t_0}$ shifted by the action of the flow
$g^t_{\mathcal{H}}$ gives again the point $\mathbf{r}(\psi_j,
\tau+t)$ on $ M^2_{t_0+t}$ but with the shifted  time
$\tau+t$. Due to definition  the coordinate $\alpha$ does not change. That is why  the coordinate $\tau$ (corresponding to $t$) is
called the {\it proper time}. Sometimes it is possible to choose $\delta$ arbitrary large, even infinity (e.g. in  the case $C=const$). But in many situation the set $\{p=
P(\psi, \tau), x=X(\psi, \tau)| \psi\in \mathbb{S},\tau\in -\infty\}$ has
the intersection points (e.g. if the trajectories  $P(\psi, \tau), x=X(\psi, \tau)$ belong to the Liouville tori), and this set  is not even the manifold. But for our purpose it is enough to work with the ``Lagrangian band'' $M^2_t$ only.
Along with the general properties of Lagrangian manifolds, the
band  $M^2_t$ has very useful additional ones. Let us present
them all for the completeness.

Let us introduce the matrices
\begin{equation*}\label{BC}
\mathcal{B}=\frac{\pa P}{\pa(t,\psi)}\equiv (\dot P,P_{\psi}),
\quad\mathcal{C}= \frac{\pa X}{\pa(t,\psi)}\equiv (\dot
X,X_{\psi})
\end{equation*}.
%The dimension of the manifold $M^2$ is equal to  2 and
It is easy
to see also that each column-vector $\begin{pmatrix}
\dot P\\ \dot X\end{pmatrix} $, $\begin{pmatrix} P_\psi\\
X_\psi\end{pmatrix} $ and  $\begin{pmatrix}
 P\\0\end{pmatrix} $ satisfies the variational system
\begin{equation}
\dot{\delta x}= \CH_{pp}\delta p +\CH_{px}\delta x,\quad
\dot{\delta p}= -(\CH_{xp}\delta p+ \CH_{xx}\delta x)
\label{variation}\
\end{equation}
Easy to check that these vectors are linear independent and obviously
two first  vectors are tangent to $M^2_t$.
\begin{lemma}(see e.g.\cite{MaslovOpMet, MaslovFed}) The following properties are true:

1)the rank of the  matrix $\begin{pmatrix} \mathcal{B}\\
\mathcal{C}\end{pmatrix} $ is equal to 2 which actually means that
dimension of $M^2_t$ is 2.

2)${}^t\mathcal{B}\mathcal{C}={}^t\mathcal{C}\mathcal{B}$ which
means that $M^2$ is Lagrangian,

3)for any positive $\varepsilon$ the matrix $\mathcal{C}-i
\varepsilon\mathcal{B}$ is not degenerate.\end{lemma}
\textsc{Proof}. The first two propositions follow from the
properties of the variational system. It is easy to check them for
$t=0$ because $\mathcal{B}=(-\nabla C(0),\mathbf{n}_\perp),
\mathcal{C}=(C(0)\mathbf{n},0)$ where
 $\mathbf{n}_\perp={}^t(-\sin\psi, \cos \psi )$. In this argument
we use the definition of the trajectories $(P,X)$, namely the
property $P|_{t=0}=\mathbf{n}(\psi), X|_{t=0}=0$, $\mathbf{n}=
{}^t(\cos\psi, \sin \psi )$. Thus according to the variational
system (\ref{variation}) the vector columns $\begin{pmatrix}
\dot P\\ \dot X\end{pmatrix} $ and $\begin{pmatrix} P_\psi\\
X_\psi\end{pmatrix} $ are linearly independent for each $t$ which
gives 1). Also a simple calculation based on the variational
system (\ref{variation}) gives that
$\frac{d}{dt}({}^t\mathcal{B}\mathcal{C}-{}^t\mathcal{C}\mathcal{B})=0$,
which gives 2). To prove 3) assume that $\mathcal{C}-i
\varepsilon\mathcal{B}$ is degenerate, then there exists a 2-D
vector $\xi\neq 0$ such that $\mathcal{C}\xi=i
\varepsilon\mathcal{B}\xi$. Consider the (complex) scalar product
0=\begin{gather*} <\xi,({}^t\mathcal{B}\mathcal{C}-
{}^t\mathcal{C}\mathcal{B})\xi>_c=<\mathcal{B}\xi,\mathcal{C}\xi>_c-
<\mathcal{C}\xi,\mathcal{B}\xi>_c=i(
\varepsilon<\mathcal{C}\xi,\mathcal{C}\xi>_c+
\frac{1}{\varepsilon}<\mathcal{B}\xi,\mathcal{B}\xi>_c)=0.
\end{gather*}
From this equation it follows that both
$\mathcal{B}\xi=0,\mathcal{C}\xi=0$ which contradicts 1). $\Box$

The same consideration allows one to obtain the following closed result.
\begin{lemma} The propositions of the previous Lemma concerning the matrices $\mathcal{B},\mathcal{C}$ is   true if one changes the matrix $\mathcal{B}$ by the matrix
\begin{equation*}\label{B'}
\widetilde{\mathcal{B}}=(\dot P-\lambda P,P_{\psi}),
\end{equation*}
where $\lambda=\langle C_x(0),\mathbf{n}(\psi)\rangle$.
\end{lemma}
Let us recall that the points $x=X(\psi^F,t)=X^F$ on $M^2_t$ where the Jacobian
$$J\equiv\rm{det}\mathcal{C}\equiv\rm{det}(\dot X,X_\psi)=0$$ are the
{\it focal} points \footnote {Note that using the Hamiltonian
system we can change $\dot X$ by $P$ in last formula as well as in
many formulas containing $\dot X$.}. Since the manifold $M^2_t$ is
generated by the curves $\Gamma_t$ as well as by the trajectories
$(P,X)$ each focal point of one of theses objects simultaneously
is a focal point for the other ones. Little later we shall show that this definition of the focal points coincides with the definition, based on equality $X_\psi=$, used in the previous sections.

Let us fix some time $t$ and consider the smooth curve
$\Gamma_t=\{p=P(\psi,t),\,x=X(\psi,t)\}$ on $M^2_t\in
\mathbb{R}^4_{p,x}$ (the ``time cut'' of $M^2_t$). Then obviously
the  front $\gamma_t=\{x=X(\psi,t)\}$ is nothing but the
projection of $\Gamma_t$ to $\mathbb{R}^2_x$. Hence the focal
points on the front are also the focal points of the manifold
$M^2_t$, and from this point of view the caustics of $M^2_t$ are
called space-time ones.

\begin{lemma} The vector-functions $\dot X$ and $X_\psi$ as well as
 vector-functions $P$ and $X_\psi$ are orthogonal:
 $\langle{\dot X,X_\psi}\rangle=\langle{P,X_\psi}\rangle=0$.
\end{lemma}
\textsc{Proof}. According to system (\ref{Ham3}) the vectors  $P$
and $\dot X$ are parallel and it is enough to prove the second
equality. Let us differentiate  $\langle{P,X_\psi}\rangle$ along
the trajectories of the system (\ref{Ham3}). We have
\begin{gather*}\frac{d}{dt}\langle{P,X_\psi}\rangle=
\langle{\dot P,X_\psi}\rangle+\langle{P,\dot X_\psi}\rangle=
\text{using\quad\ref{Ham4}}=
-|P|\langle{C_x,X_\psi}\rangle+\frac{C^2}{C_0}\langle{P,P_\psi}\rangle
+\frac{\pa C^2}{\pa
\psi}\frac{1}{C_0}\langle{P,P}\rangle=\\-|P|\frac{\pa C}{\pa
\psi}+ \frac{1}{2 C_0}\frac{\pa (C^2 P^2)}{\pa \psi}
+\frac{C|P|}{C_0}|P|\frac{\pa C}{\pa \psi}=-|P|\frac{\pa C}{\pa
\psi}+ \frac{1}{2 C_0}\frac{\pa (C_0^2)}{\pa \psi} +|P|\frac{\pa
C}{\pa \psi}=0.\end{gather*} But $X|_{t=0}=0$, thus
$\langle{P,X_\psi}\rangle|_{t=0}=0$ and Lemma is proved. $\Box$
\begin{cor} 1) The following equality is true
$J=\rm{det}(\dot X,X_\psi)=\pm|\dot X|\,|X_\psi|$.
2) The point $x=X(\psi,t)$ on the front $\gamma_t$, or the point
$\mathbf{r}=(p=P(\psi,t),x=X(\psi,t)$ on $\Gamma_t\in M^2_t$ is a focal one
if and only if  $J=\rm{det}(\dot X,X_\psi)=0$.
\end{cor}
According to the equality $|\dot X|=C(X)\,\,$ $J=\rm{det}(\dot
X,X_\psi)$ as well as the Jacobian  in some  neighborhood of
$\gamma_t$ can be equal to zero if and only if $X_\psi=0$. Thus
the last equation really determines the focal points from the
point of view of the Lagrangian manifold also.
\begin{lemma}  In the focal point $x=x^F=X(\psi^F,t)$
1) $\langle P^F,P_{\psi}^F\rangle=0$, but 2) $P_{\psi}^F\neq 0$,
3) $\frac{d J}{dt}=\frac{C^2_F}{C_0}\rm{det}(\dot X^F,P_\psi^F)$, where as it was before $C_0=C(0)$ and $C_F=C(X^F).$
\end{lemma}
\textsc{Proof}. According to the conservation law (\ref{Ham4})
$\langle P,P_{\psi}\rangle(\psi^F,t)=\langle
\nabla(\frac{C_0^2}{C^2(x)}), X_{\psi}\rangle(\psi^F,t)=0$. To
prove the second inequality one can mention that the
vector-function $(P_{\psi},X_{\psi})^T$ satisfies the linear
(variational) system with non-zero initial condition. Thus both
components of the solution cannot be equal to zero. To prove 3) we
write $\frac{d J}{dt}|_{\psi=\psi^F}=
[\rm{det}(\dot X,\dot X_\psi)+\rm{det}(\ddot X,X_\psi)]_{\psi=\psi^F}=
\text{using\quad\ref{Ham4}}=$\break$\rm{det}(\dot X,P <\nabla\frac{C^2(X)}{C_0},
X_\psi>)|_{\psi=\psi^F}+$$
\rm{det}(\dot X,(\frac{P_\psi C^2(X)}{C_0}))|_{\psi=\psi^F}=\frac{C^2_F(X)}{C_0}\rm{det}(\dot X,P_\psi)|_{\psi=\psi^F}$. $\Box$
\begin{cor} In the focal point
1) $\frac{d J}{dt}=\frac{C^2_F}{C_0}\rm{det}(\dot
X,P_\psi)=\pm|\frac{C^2(X^F}{C_0}\dot X|\,|P_\psi|(\psi^F,t)\neq
0$;$\quad$ 2) during the passage through the focal point the
Jacobian $J$ changes its sign from - to + if $\rm{det}(\dot
X,P_\psi)|_{\psi=\psi^F}>0$ and from + to - if $\rm{det}(\dot
X,P_\psi)|_{\psi=\psi^F}<0$;$\quad$ 3) There exists $t_{cr}$ such
that $J(\psi,t)>0$ for $ 0<t<t_{cr}$.
\end{cor}
\textsc{Proof}. To prove 3) it is enough to note that
$\rm{det}(\dot X, P_\psi)|_{t=0}=C(0) \rm{det}(\mathbf{n(\psi)},
\mathbf{n(\psi)}_\perp)= C(0)$. $\Box$

\subsection{The Maslov and Morse indices.$\label{maslovmorse}$}

 As we said before
the front $\gamma_t$ as well as the curve $\Gamma_t$ is
partitioned into arcs with the focal points at their ends and it
is formed by the ends of trajectories having the same topological
structure. This means that they have similar crossing (on $M^2_t$ )
with the focal points and the same topological characteristic,
i.e. the Maslov index. But it coincides with the the Morse index for the considered situation (see subsection (\ref{fronts})).
Let us prove this proposition.
%{\bf NNEW 
Let us remind some necessary definitions and constructions. It is
needless to say that there exist different definitions of the
Maslov index. The original definition \cite{MaslovTeorVozm} is based on calculation of indices of inertia of matrices $\frac{\pa(x_1,p_2)}{\pa(x_1,x_2)}|_{M^2_t}$, $\frac{\pa(p_1,x_2)}{\pa(x_1,x_2)}|_{M^2_t}$,
$\frac{\pa(p_1,x_2)}{\pa(x_1,p_2)}|_{M^2_t}$  etc. It is not very convenient
in practical calculation. Thus we want to present below one
\cite{MaslovOpMet,Arnold,DobZh} which, from our point of view, is
more progmatic for computer calculations.% ENDNNEW}
We already pointed out
that the Maslov index of the points on $x\in \gamma_t$ is the
index of the {\it nonsingular} point
$\mathbf{r}(\psi,t)=(P(\psi,t),X(\psi,t)$ on the Lagrangian
band $M^2_t$.  According to the procedure from \cite{Maslov,
MaslovFed,Arnold, DobrZhev} one needs to fix the index $m^0$ in
some marked nonsingular point $p=P(\psi_0,\zeta),x=X(\psi_0,\zeta)$ on
$M^2_0$ and then to find the change of the argument of the determinant
of the $2\times2$ matrix $\mathbb{C}_\varepsilon^{(1,2)}=
\big(\mathcal{C}- i\varepsilon \mathcal{B}\big)\equiv(\dot X-i
\varepsilon \dot P, X_\psi-i \varepsilon  P_\psi\big)$ along one of the paths described below   joining the marked
point $p=P(\psi_0,\zeta),\quad x=X(\psi_0,\zeta)$ with the given
nonsingular point $p=P(\psi,t),x=X(\psi,t)$, more precisely
\begin{gather}
m(\psi,t)=m(\psi_0,t_0)+\Delta m,\quad \Delta m=
\frac{1}{\pi}\lim_{\varepsilon\rightarrow+0}\rm{Arg}\,\rm{det}\big(\dot
X-i \varepsilon \dot P, X_\psi-i \varepsilon P_\psi\big)(\psi,
t)|^{\psi,t}_{\psi_0,t_0}.\label{4.3}
\end{gather}
From  definition
\eqref{4.3} it follows the fact that we used before: the index can
change (jump) only crossing a focal point. In fact, if the point
$(\psi, t_\psi)$ is a regular point then $det(\dot X, X_\psi)$ is
different from zero so the increment of the argument of the
determinant goes to zero when $\varepsilon$ goes to zero,
otherwise, if the determinant of $(\dot X, X_\psi)=0$, as it
happens in a focal point, then the increment of the determinant in
\eqref{4.3} is different from zero when $\varepsilon$ goes to
zero.
We know  (see Corollary from  Lemma 4) that  the Jacobian $J=\det\mathcal{C}(\psi,\zeta)>0$ for    small enough positive $\zeta$. Thus  all the points on the front
$\gamma_\zeta$ are nonsingular. So we choose one of the point $(P(\psi_0,\zeta),X(\psi_0,\zeta))$ and put Maslov index $m(\psi_0,\zeta)=0$.
It is possible (and natural) to use one of the  of the following
paths. 1) To  move first along the trajectory
$P(\psi_0,\tau),X(\psi_0,\tau)$ starting from $\tau=\zeta$ until
$\tau=t$, then to move from the point $P(\psi_0,t),X(\psi_0,t)$
along the curve  $\Gamma_t$ to the point
$P(\psi,t),X(\psi,t)$ changing the angle $\psi$. As we will see a
little later this choice is not very convenient from the point of
view of computer realization. The choice 2) is to move first from the
point with the angle $\psi_0$ to the point with the angle $\psi$
along the curve $\Gamma_{\zeta}$, and then to
move along the trajectory with the angle $\psi$, changing time
from $\zeta$ to $t$. %{\bf NEW till the end of subsection}
It could happen that during the motion along some closed path on  $M^2_t$ one can get nontrivial increment of argument of the determinant of the matrix $\mathbb{C}_\varepsilon^{(1,2)}=
\big(\mathcal{C}- i\varepsilon \mathcal{B}\big)\equiv(\dot X-i
\varepsilon \dot P, X_\psi-i \varepsilon  P_\psi\big)$ and nontrivial Maslov index of this closed path. The following lemma shows that it is not so.

\begin{lemma} The increment of argument of the determinant of the matrix $\mathbb{C}_\varepsilon^{(1,2)}=
\big(\mathcal{C}- i\varepsilon \mathcal{B}\big)\equiv(\dot X-i
\varepsilon \dot P, X_\psi-i \varepsilon  P_\psi\big)$ along any closed path on $M^2_t$ is equal to zero. Thus  the Maslov index of any closed path on $M^2_t$ is also equal to zero.
\end{lemma}
\textsc{Proof} First let us show that Lemma is true for $t=0$. Obviously all closed paths on $M^2_0$ are homotopic one to another. Let us choose as a path the curve $\Gamma_\zeta$, $\delta>\zeta>0$. According to Corollary from  Lemma 4  the Jacobian $J=\det\mathcal{C}(\psi,\zeta)>0$ for   $\zeta$ small enough. Thus the increment of argument of determinant of matrix $\mathcal{C}-i\varepsilon \mathcal{C}$ over $\Gamma_\zeta$ is zero and Maslov index of this path is also zero. Hence according to the property of Maslov index it is equal to zero for any other closed path on $M^2_0$. Any path on $M^2_t$ could be obtain from some path from $M^2_0$ by means of the canonical transform (the flow) $g^t_{\mathcal{H}}$. But this transform preserves the Maslov index of closed path. $ \Box$
\begin{rem}  Thanks to this statement   the Bohr-Sommerfeld quantization rule does not appear in asymptotic solutions, although the paths $\Gamma_t$ on $M^2_t$ are not contractible. This fact  is also natural for Lagrangian manifolds associated with the asymptotics of Green functions for evolution equations in Euclidian space (like van Vleck formula for the nonstationary Schr\"odinger equation).
\end{rem}
Taking into account this Lemma, choosing the the second way for the calculation of the Maslov index and putting  $m(\psi_0,\zeta)=0$ for some small positive $\zeta<\delta$ and some fixed fixed $\psi_0$, (and hence for {\it any} $\psi$)  we can write
for the index
$m(\psi,t)$:
\begin{gather}
m(\psi,t)=\Delta m,\quad \Delta m=
\frac{1}{\pi}\lim_{\varepsilon\rightarrow+0}\rm{Arg}\,\rm{det}\big(\dot
X-i \varepsilon \dot P, X_\psi-i \varepsilon P_\psi\big)(\psi,
t)|^{\psi,t}_{\psi,\zeta}.\label{4.3a}
\end{gather}
 The Corollary of Lemma 3 allows us (in the considered
problem) to simplify the definition \eqref{4.3}. Let us analyze
the determinant
\begin{gather}
\rm{det}\big(\dot X-i \varepsilon \dot P, X_\psi-i \varepsilon
P_\psi\big)(t,\psi)= \rm{det}\big(\dot X,X_\psi\big)-i \varepsilon
\rm{det}\big(\dot X,P_\psi\big)-i \varepsilon \rm{det} \big(\dot
P,X_\psi\big)-\varepsilon^2\rm{det}\big(\dot
P,P_\psi\big).\label{arg}
\end{gather}

Since $\rm{det}\big(\dot X,P_\psi\big)$ is not equal to zero and
the term $\rm{det} \big(\dot P,X_\psi\big)=0$ in the focal point,
then the third term of equation \eqref{arg} can be omitted.
Similarly taking into account the fact that the jump of the index
is an integer number it easy to show that the term
$\varepsilon^2\rm{det}\big(\dot P,P_\psi\big)$ does not play any
role in the calculation of $\Delta m$ and also can be omitted.
Thus instead of the determinant $\rm{det}\big(\dot X-i \varepsilon
\dot P, X_\psi-i \varepsilon  P_\psi\big)(t,\psi)$ we can use the
determinant $\rm{det}\big(\dot X, X_\psi-i \varepsilon
P_\psi\big)(\psi, t)$.

Thus one needs to found the jumps of $m$ during
the crossing of the focal points and find
\begin{gather}
\Delta m= \frac{1}{\pi}\lim_{\varepsilon\rightarrow+0}\rm{Arg}\,
\rm{det}\big(\dot X, X_\psi-i \varepsilon  P_\psi\big)(\psi,\tau)|
^{\tau=t_\psi+\Delta t}_{\tau=t_\psi-\Delta t},\label{3.4b}
\end{gather}
\noindent where $\Delta t>0$ is small enough, and $t_\psi$ is the
time at which the trajectory crosses the focal point (with
coordinates $\psi,t_\psi$)with the angle $\psi$. There may exist
several such $t_\psi$, but all of them, according to the point $1$
of the Corollary of Lemma 3, are isolated with respect to $t$. In
fact the derivative of the Jacobian is different from $0$ in a
focal point so the Jacobian has an isolated zero and so those
zeros cannot accumulate in one point. Now we use again this
Corollary. Obviously we can change $\tau$ with $t^F$ in the right
hand side of \eqref{3.4b}. But the term $\rm{det}\big(\dot
X,P_\psi\big)(\psi^F,t^F)$ characterizes the increasing or
decreasing of the first term. Hence if $\rm{det}\big(\dot
X,P_\psi\big)(\psi^F,t^F)<0$, then the argument of the complex
vector in \eqref{3.4b} changes on the upper half plane from
$O(\varepsilon)$ to $\pi +O(\varepsilon)$ and $\Delta m=1$.  If
$\rm{det}\big(\dot X,P_\psi\big)(\psi^F,t^F)>0$ then the argument
of the vector in \eqref{3.4b} changes from $\pi+O(\varepsilon)$ to
$2\pi +O(\varepsilon)$ and {\it again} $\Delta m=1$. Thus we
obtain the following important result.

\begin{lemma} The Maslov index $m(\psi,t)$ of any nonsingular point
$(p=P(\psi,t),x=X(\psi,t))\in \Gamma_t$ with the projection
$x=X(\psi,t)$ on the front is equal to number of focal points on
the trajectory $P(\psi,\zeta),X(\psi,\zeta), \zeta\in(+0,t)$, i.e. it
coincides with the Morse index of this trajectory.
\end{lemma}

\subsection{The behavior of the front near the focal points.$\label{focal}$}
%{\bf NEW} Here we change $\mu$ by $q$!!!!!
For the future developments it is useful to have the description
of the wave front $\gamma_t$ in a neighborhood of the focal
points. The focal points which are ends of arcs $\gamma_t^j$ of
the wavefront belong to the {\it caustics}, a well known concept
in geometrical optics and in the space-time wave theory. It is an
important fact that in our problem the caustics do not depend on
the time because the family of manifolds (bands) $M^2_t$ is
invariant with respect to
the phase shift $g_{\mathcal{H}}^t$. One may distinguish two types
of sets organized by the focal points with stable and unstable
structures with respect to small changes of the Lagrangian
manifold $M^2_t$ (or of functions  $X(\psi,t), P(\psi,t)$). The
first type is under the so-called ``general position'', there
exist only a finite numbers of them and in the considered 2-D
situation namely there are only two types: the so-called {\it
fold} and {\it cusp}\cite{Arnold}. These curves are presented in
the Fig.1. Sometimes there exist also different focal sets with
unstable structure. For instance the circle
$\Gamma_0={p=\mathbf{n}(\psi), x=0}$ on $M^2_0$, has the point $x=0$
as the projection $\Gamma_0$ from $M^2_T$ (or $\mathbb{R}^4_{p,x}$)
to $\mathbb{R}^2_x$. Rotating  a little the coordinate system  in
$\mathbb{R}^4_{p,x}$ one can obtain a small ellipse on
$\mathbb{R}^2_x$ instead of the point $x=0$. But we show below
that namely  this ``unstable singular'' circle determine localized
functions in the asymptotic constructions.

Taking into account the smoothness of the vector-functions
$X(\psi,t)$ one can easily describe the behavior of the wavefront
near the focal points. Namely let us fix the time $t$ and let
$\psi^F=\psi^F(t)$ define the angle (coordinate on $\Gamma_t$) of
the focal point $X^F=X(\psi^F,t)$. We put $y=\Delta
\psi=\psi-\psi^F$. Let $n\geq 2$ be the minimum degree of the
Taylor expansion of the function $X$ around $\psi$ with increment
$y$. We say that the focal point $X^F$ is not completely
degenerate if $n\neq\infty$. From the point of view of this
definition the point $p=P_0\equiv\mathbf{n}(\psi), x=X_0=0$ is a
complete degenerate one.

Shifting the origin into the focal point and rotating the
coordinates we can ``kill'' the second component of the $n$--th
derivative vector  $X_{\psi}^{(n)}(\psi^F(t),t)$ and write $x'_1=
a y^n+O(y^{n+1})$, $x'_2= b y^k+O(y^{k+1})$. Here $a\neq 0$,
$b\neq 0$ are Taylor coefficients, the integer $k>n$ and the prime
indicates the new coordinates; actually $a$ and $b$ depend on the
time $t$, but now for us it is not important. It is clear that the
previous Lemmas are true also in the new coordinates.

\begin{lemma} In the non degenerate case only one opportunity is possible: $k=n+1$.
\end{lemma}
\textsc{Proof}. In new coordinates
\begin{gather*}X'=\begin{pmatrix}
a y^n+O(y^{n+1})) \\
b y^k+O(y^{k+1})
\end{pmatrix},\quad X'_{\psi}=\begin{pmatrix}
a n y^{n-1}+O(y^n) \\
b k y^{k-1}+O(y^k)
\end{pmatrix},
\end{gather*}
Since the vector $P'$ is orthogonal to $X'_{\psi}$ everywhere,
and, according  to the conservation of the Hamiltonian $\CH$,
$P'\neq 0$, then we can write
\begin{gather*}P'=q \begin{pmatrix}
-b k y^{k-n}+O(y^{k-n+1})) \\
a n +\tilde a y +O(y^{2})\end{pmatrix} \text{and}\qquad P'_{\psi}=q \begin{pmatrix}
-b k(k-n) y^{k-n-1}+O(y^{k-n})) \\
\tilde a  +O(y)
\end{pmatrix},
\end{gather*}
where the factor $q(t)\neq 0$, $q\tilde a(t)$ is proportional
to the Taylor coefficient just after $q a(t)$. Taking into
account point 1) of Lemma 4 we immediately find $\tilde a(t)=0$. But from this
it follows that if $n,k\neq \infty$ and  $k>n+1$ then in the focal
point $P'_{\psi}=0$ which contradicts point 2) of Lemma 4. $\Box$.
\begin{cor} Let $n\neq \infty$ then in the neighborhood of the
focal point $x=X^F$ in new coordinates
\begin{gather}\nonumber X'=\begin{pmatrix}
a y^n+O(y^{n+1})) \\
b y^{n+1}+O(y^{n+2})
\end{pmatrix},\quad X'_{\psi}=\begin{pmatrix}
a n y^{n-1}+O(y^n) \\
b (n+1) y^{n}+O(y^{n+1})
\end{pmatrix},\\
P'=q \begin{pmatrix}
-b (n+1) y+O(y^{2})) \\
a n  +O(y^{2})\end{pmatrix}, \qquad P'_{\psi}=q \begin{pmatrix}
-b (n+1) +O(y)) \\
O(y)
\end{pmatrix},\label{3.4a}
\end{gather} and
\begin{equation} \rm{det}(\dot X,X_\psi)=-\frac{q a n}{|q|}C(X^F) y^{n-1}+O(y^n),\qquad
\rm{det}(\dot X,P_\psi)=|q| b( n+1)
C_F+O(y),\label{3.4}\end{equation}
where it was before $C_F=C(X^F)$. The last two equalities do
not depend on the choice of the coordinates.
\end{cor}
 Thus in agreement with the Lemma 3  the
determinant $\rm{det}(\dot X,P_\psi)$ does not change its sign in
a neighborhood of a non degenerate focal point.

Finally in the non completely  degenerate case, in the neighborhood
of the focal point, we have $X'_1=a y^n+O(y^{n+1}),\, X'_2=b
y^{n+1}+O(y^{n+2}),\,P'_1=-qb(n+1) y+O(y^2),\,P'_2=q a n +O(y^2)$.

%{\bf NEW} %Without loss of generality we can assume for a moment that$a>0$ for odd $n$ and that $b>0$ for even $n$.{\bf ENDNEW}

 Omitting the higher
corrections we find the equation for the part of the front
$\gamma_t$ in the neighborhood of the focal point $x^F$: $x'_1=a
y^n,\, x'_2=b y^{n+1}$. The sign of $ab$  for odd $n$ and the sign
of $a$  for even  $n$ defines the direction of the passage from
higher to lower leaves for odd $n$ and from left to right leaves for
even  $n$, see Fig.5. Let us note also that in the general case $n$ is equal
to 2 or 3 only \cite{Arnold1,Arnold2}.
%{\bf HERE is Fig. 5} {\bf NEW}

It is convenient to express the coefficients $a,b,q $ via $P,X$ and their derivatives in the focal point $\psi^F(t)$.
 Putting in formulas \eqref{3.4a}
$y=0$ we find
$$P'_1=0,\qquad P'_2= q a n,\qquad P'_{1\psi}=-bq (n+1),\,\,\text{for}\,\,\, \psi= \psi^F(t),$$
and
$$a=\frac{{X'_1}^{(n)}}{n!}\equiv\frac{1}{n!}\frac{\pa^n X'_1}{\pa \psi^n},\,\,\,\,
b=-\frac{P'_{1 \psi} {X'_1}^{(n)}}{(n-1)!(n+1)P'_2},\,\,\,\,
q=\frac{P'_2 (n-1)!}{{X'_1}^{(n)}}\,\,\,\text{for}\,\,\, \psi= \psi^F(t).$$
The directions of the vectors $P$ and $\dot X$ coincide in the focal points. Thus we see that the coordinates with index ``prime'' could be chosen the  coordinates introduced in \eqref{change}. This gives:
$P'_2=|P^F|\equiv\frac{C_0}{C_F},\quad$ $P'_1=\langle \mathbf{k}_1,P_\psi^F\rangle\equiv-\det(\dot X^F,P_\psi^F)/C_F\equiv-\tilde J_F/C_F,\quad$
${X'_1}^{(n)}=\langle \mathbf{k}_1,{X}^{(n)F}\rangle\equiv-\det(\dot X^F,{X}^{(n)F})/C_F\equiv-J^{(n)}_F/C_F\quad$ and
\begin{equation}\label{coef}
a=-\frac{J^{(n)}_F}{n!C_F},\,\,\,\,
b=-\frac{n\tilde J_F J^{(n)}_F}{(n+1)!C_F C_0},\,\,\,\,
q=-\frac{C_0(n-1)!}{J^{(n)}_F}.
\end{equation}

\subsection{The jumps of the Maslov index along the front.}

Let us find the jumps $\Delta m$ of the Maslov index during the
passage through the focal points along the front. We fix the time
$t>0$ and consider the path to cross non degenerate focal points
(studied above) starting from the angle $\psi_0-\delta$ and ending
at the angle $\psi_0+\delta$.

\begin{lemma} The following equalities are true (see Fig. 3): for odd
$n$ $\Delta m=0$ %(Fig.3), 
for even $n$ and 
$\rm{sign}(\tilde J_F J^{(n)}_F)=\pm 1\,\, \Delta m=\pm 1 $. 
%(Fig.3b,c).
\end{lemma}
\textsc{Proof}. Similarly to the proof of Lemma 4, taking into
account the inequality $\rm{det}\big(\dot X,P_\psi\big) \neq 0$
instead of \eqref{4.3} we can write:
 \begin{gather}\Delta m= \frac{1}{\pi}\lim_{\varepsilon\rightarrow+0}\rm{Arg}\,
 \rm{det}\big(\dot X, X_\psi-i \varepsilon  P_\psi\big)(\psi,t)|^{\psi_0+\delta,t}_{\psi_0-\delta,t}
 =\\
\frac{1}{\pi}\lim_{\varepsilon\rightarrow+0}\rm{Arg}\,
[ -q a C(X^F) y^{n-1}+O(y^n)-i \varepsilon( q^2 C(X^F)b(n+1)+O(y))]|^{y=\delta}_{y=-\delta}=\\
\frac{1}{\pi}\lim_{\varepsilon\rightarrow+0}\rm{Arg}\,
[ -   y^{n-1}-i \varepsilon( \frac{q b(n+1)}{a})]|^{y=\delta}_{y=-\delta}.\label{3.4a}
\end{gather}
We see now that the complex vector-function $ -y^{n-1}-i
\varepsilon( \frac{q b(n+1)}{a})$ lies in one half plane for even values
of $n$ for each $y$ and in one quadrant for odd values of n. So
for $n$ even one has $\Delta m =-1$ if $qab>0$ and $\Delta m=1$
if $qab<0$ while for $n$ odd $\Delta m=0$. To finish the proof it is enough to  take into account formulas \eqref{coef}. $\Box$

 Coming back to the original variables we can make the following conclusion.

 \begin{lemma}During  the motion along the front $\gamma_t$

1) the Maslov index does not change if the path does not cross the
focal points or if the Jacobian $J=\rm{det}\,(\dot X,X_{\psi})$
does not change the sign after the passage through the focal
point;

2)let the Jacobian $J=\rm{det}\,(\dot X,X_{\psi})$  change sign
after the passage through the focal point then $\Delta m=1$ if
the signs of $J=\rm{det}\,(\dot X,X_{\psi})$ and $\tilde
J=\rm{det}\,(\dot X,P_{\psi})$ coincide in the end of the path and
$\Delta m=-1$ if the signs of $J=\rm{det}\,(\dot X,X_{\psi})$ and
$\tilde J=\rm{det}\,(\dot X,P_{\psi})$ are different.
\end{lemma}

\begin{exam} $\label{bank3}$

Let us check the correspondence of this
conclusion with the index $m(x\in\gamma_t^2)$ for the example with
the axial symmetric bank. Obviously $J>0$ on the arc $\gamma_t^1$.
Since the trajectories coming to arc $\gamma_t^2$ meet the focal
point only one time and $\dot J\neq 0$ in this point, then $J<0$
on the arc $\gamma_t^2$. We have shown in the Example
\eqref{bank2} that $\tilde J<0$ in the focal point.
%Let us find the sign of $\tilde J$ in the focalpoint. To this end
%differentiate the integrals of motion\eqref{4.2} of Hamiltonian system with respect to the angle $\psi$
%and put $\psi= \psi_{F}(t)$.  We find
%$$\langle P,P_\psi \rangle
%=0,\qquad X_2P_{1\psi}-X_1P_{2\psi}=\rho_0\sin
%\psi_{F}(t)\qquad\text{for}\quad \psi= \psi_{F}(t).$$
%The solution of this equation is $$\begin{pmatrix}P_{1\psi}\\P_{2\psi}\end{pmatrix}=
%\frac{\rho_{0}\sin \psi_{\rm {cr}(t)}}{\langle P,X\rangle}
%(\psi_{\rm {cr}(t)},t)\begin{pmatrix}P_{2}\\-P_{1}\end{pmatrix}$$ and
%\frac{|P|}{C(X^F)}\rm{det}\big(\dot
%X,P_{\psi}\big)=\rm{det}\big(P,P_{\psi}\big)=
%-\frac{P^2\rho_0}{\langle P,X\rangle}\sin \psi_{\rm {cr}(t)}.$ Obviously the angle
%l$\psi^F(t)$ belongs to the interval $(0,\pi)$ and ${\langle P,X\rangle}>0$.

Thus $\Delta m=1 $ if we move from the arc $\gamma_t^1$ to the arc
$\gamma_t^2$ %{\bf but if the signs of $J$ and $\tilde J$ coincide then $\Delta m$ should be $-1$ SERGEY: there was a mistake in Lemma, I corrected it}
and $\Delta m=-1$ if we move in the
opposite direction. Finally we find $m(x\in\gamma_t^2)=1$, which
agrees with our previous conclusion. In the critical moment of
time when the first focal point just appears the Jacobian
$J(\psi,t_{\rm{cr}})$ does not change the sign after crossing the
focal point. Thus the index $m$ also does not change and equals to
$0$ in all regular points of the front $\gamma^{t_{\rm{cr}}}$. So
we see the difference in the jump of the Maslov index depends on
the direction of crossing the focal points on the manifold $M^2_t$.
\end{exam}

\subsection{ Canonical planes in the phase space, 
nonsingular and singular  maps.}
To construct the asymptotic solution of problem  
\eqref{WEq}, \eqref{WEq1}
in the neighborhood of the focal points we need 
additional construction, related with the fronts, 
maps covering  Lagrangian bands $M^2_t$, indices of 
these maps etc. Let us describe 
them also briefly, using notation introduced in \cite{DobrZh}.

The 2-D planes   with the {\it focal} coordinates
$x^{(1,2)}=(x_1,x_2)$, $x^{(1,0)}= (x_1,p_2)$,
$x^{(0,2)}=(p_1,x_2)$, $x^{(0,0)}=(p_1,p_2)$ in the phase space
$\mathbb{R}^4_{p,x}$ are called symplectic canonical planes. It is
convenient to introduce the multi-indices $I=(1,2)$ corresponds to
the canonical plane $(x_1, x_2)$, $I=(1,0)$ to $(x_1,p_2)$,
$I=(0,2)$ to $(p_1,x_2)$, $I=(0,0)$ to $(p_1,p_2)$. \footnote
{These multi-indices indicate the replacement of the coordinate
$x_j$ corresponding to the entry zero of the pair $(a_1, a_2$ by
the momentum $p_j$ ({\it with the same number $j$}).} We denote
also $p^{(1,2)}=(p_1, p_2)$, $p^{(1,0)}= (p_1,-x_2)$,
$p^{(0,2)}=(-x_1,p_2)$, $p^{(0,0)}=(-x_1,-x_2)$. We call $I$ the
index of singularity. It is convenient to mark the canonical plane
by the corresponding index $I$ and write $\mathcal{R}^2_I$.

According to general property of Lagrangian manifold one can cover $M^2_t$ %and $\Lambda^2$
by the maps $\Omega^{I_j}_j$ with the numbers $j$ such that there
exist one-to-one map from $\Omega^{I_j}_j$ to its projection to
the canonical plane $\mathbb{R}^2_{I_j}$. This means the
following. Along with the matrices
$\mathcal{B}^{(1,2)}=\mathcal{B}$,
$\mathcal{C}^{(1,2)}=\mathcal{C}$ it is convenient
to introduce the matrices
\begin{gather}
\mathcal{B}^{(0,2)}(\psi,\tau)=\begin{pmatrix}\nonumber
-\frac{\partial X_1}{\partial t}&-\frac{\partial X_1}
{\partial\psi} \\
\frac{\partial P_2}{\partial t}& \frac{\partial P_2}{\partial
\psi}
\end{pmatrix},\,\,\,\mathcal{C}^{(0,2)}(\psi,\tau)=\begin{pmatrix}
\frac{\partial P_1}{\partial t}&\frac{\partial P_1}
{\partial\psi} \\
\frac{\partial X_2}{\partial t}& \frac{\partial X_2}{\partial
\psi}
\end{pmatrix},\\ \mathcal{B}^{(1,0)}(\psi,\tau)=\begin{pmatrix}
\frac{\partial P_1}{\partial t}&\frac{\partial P_1}
{\partial\psi}\\
-\frac{\partial X_2}{\partial t}& -\frac{\partial X_2}{\partial
\psi}\end{pmatrix} ,\,\,\,
\mathcal{C}^{(1,0)}(\psi,\tau)=\begin{pmatrix} \frac{\partial
X_1}{\partial t}&\frac{\partial X_1}
{\partial\psi}\\
\frac{\partial P_2}{\partial t}& \frac{\partial P_2}{\partial
\psi}.\label{matr}
\end{pmatrix}
\end{gather}
\begin{gather} \mathcal{B}^{(0,0)}(\psi,\tau)=-\mathcal{C}=\begin{pmatrix}
-\frac{\partial
X_1}{\partial t}&-\frac{\partial X_1}
{\partial\psi}\\
-\frac{\partial X_2}{\partial t}& -\frac{\partial X_2}{\partial
\psi}\end{pmatrix} ,\,\,\,
\mathcal{C}^{(0,0)}(\psi,\tau)
=\mathcal{B}
= \begin{pmatrix} \frac{\partial P_1}{\partial t}&\frac{\partial P_1}
{\partial\psi}\\
\frac{\partial P_2}{\partial t}& \frac{\partial P_2}{\partial
\psi}.\label{matr}
\end{pmatrix}
\end{gather}
The matrices $\mathcal{C}^I$ give the Jacobians
$J^I(\psi,\tau)=\det C^I.$ Then in each map $\Omega^{I_j}_j(\tau)$
$J^{I_j}\neq 0.$ The maps with the indices $I_j=(1,2)$ are
nonsingular  ones, all others are singular ones with the focal
coordinates $x^{I_j}$. Note that {\it for practical application
sometimes it useful to choose some rotated coordinates
$(x'_1,x'_2)$ and $(p'_1,p'_2)$ in some maps $\Omega^{I_j}_j$}. It
is important to remember, that {\it the Jacobians $J=J^{(1,2)}$
and $J^{(0,0)}$ are invariant with respect to the rotation, but
the Jacobians $J^{(0,2)},J^{(1,0)}$ are not}. Actually in the
considered problem the maps with index $I=(0,0)$ are not needed.

\begin{lemma} For any time $t$ there exists a finite covering of
the neighborhood of $\Gamma_t$ from Lagrangian manifold $M^2_t$ by
the maps $\Omega^{I_j}_j$ with the indices
$I_j=(1,2),\,I_j=(1,0),\,I_j=(0,2)$.
\end{lemma}
\textsc{Proof}. It is enough to prove that at least one  of the
Jacobians $J^{(0,2)}$ or $J^{(1,0)}$ is not equal to zero in each
focal point.  Assume that both $J^{(0,2)}=0$ and $J^{(1,0)}=0$ in
the point $\psi^F,t^F$. Since $X_\psi=0$ in the focal point, this
means that $\dot X_1 P_{2\psi}=\dot X_2 P_{1\psi}=0$ and $\det
(\dot X,P_{\psi}$=0. But according to Lemmas 2,3 the vector
$P_\psi\neq 0 $ in the focal point and  $P_\psi $ is orthogonal to
$\dot X $, which is nonzero everywhere. This contradiction proves
this lemma. $\Box$

\begin{rem}

1) It is possible  to prove a similar proposition about the whole
manifold $M^2_T$, but, since $M^2_T$ is noncompact, the number of maps
is infinite, and we do not need this fact. 2) For practical
applications sometimes it useful to choose rotated coordinates
$(x'_1,x'_2)$ and $(p'_1,p'_2)$ in some maps $\Omega^{I_j}_j$. For
instance in the neighborhood of non degenerate focal points
$X(\psi^F(t),t)$ in the situation considered in subsection
\eqref{focal} it is convenient to choose the axis  $x'_2$
coinciding with the vector $\dot X(\psi^F(t),t)$  (see
subsection{focal}).
\end{rem}

\begin{exam}$\label{Gamma0}$

1) Construct  first the covering of the neighborhood of the
``initial'' curve $\Gamma_0=\{p=\mathbf{n}(\psi),x=0\}$ on the
Lagrangian manifold $M^2_t$. The projection of $\Gamma_0$ from the
phase space $\mathbb{R}^4_{px}$ on the space $\mathbb{R}^2_{x}$ is
shrunk to the point $x=0$, in this point $X_\psi=0$ for each
$\psi$. Thus according to our classification the point $x=0$ is a
completely degenerate point of the ``front'' $\gamma_0$. Using the
definition of $P(\tau,\psi), X(\tau,\psi)$ as solution of the
system \eqref{Ham3} we find for $\tau=t=0$: $$\dot
P=-\mathbf{C}'_{0}\equiv -\nabla C(x)|_{x=0},\, \dot
X=\mathbf{n}(\psi)C(x),\,
P_\psi=\mathbf{n}_\perp(\psi),\,X_\psi=0,$$ where
$\mathbf{C}'_{0}\equiv -\nabla C(x)|_{x=0}$ is a vector with
components $C_{10}=\frac{\pa C}{\pa x_1}(0),\,\, $
$C_{20}=\frac{\pa C}{\pa x_2}(0).$ Hence
$J^{(0,0)}|_{\Gamma_0}=J|_{\tau=0}=\det(\dot X,
X_{\psi})|_{\tau=0}=0.$   But \begin{gather}\label{jac2}
J^{(1,0)}|_{\Gamma_0}=\det\begin{pmatrix}
C_0\cos\psi&0\\
-C_{02}&\cos\psi
\end{pmatrix}=
C_0\cos^2\psi,\quad J^{(0,2)}|_{\Gamma_0}=\det\begin{pmatrix}
-C_{10}&-\sin \psi\\
C_0\sin\psi&0
\end{pmatrix}
=C_0\sin^2\psi.
 \end{gather}
Thus the neighborhood of the curve $\Gamma_0$ can be covered by
four singular maps (see Fig....): the map $\Omega^{(1,0)}_1$,
covering the arc with the angle
$-\frac{\pi}{4}-\zeta<\psi<\frac{\pi}{4}+\zeta, \quad$ the map
$\Omega^{(0,2)}_2$, covering the arc  with the angle
$\frac{\pi}{4}-\zeta<\psi<\frac{3\pi}{4}+\zeta, \quad$the map
$\Omega^{(1,0)}_3$, covering the arc  with the angle
$\frac{3\pi}{4}-\zeta<\psi<\frac{5\pi}{4}+\zeta \quad$ and the map
$\Omega^{(0,2)}_1$, covering the arc with the angle
$\frac{5\pi}{4}-\zeta<\psi<\frac{7\pi}{4}+\zeta $. Here and
furthermore  in these examples $\zeta$ is a small enough positive
number. The choice of these maps is not unique but the final
results does not depend on the particular choice it is necessary
that they are $4$.

2) For  $t$ less than $t_{\rm{cr}}$ one can cover the neighborhood
of $\Gamma_{t}$ by two nonsingular  maps (see Fig....) e.g.
$\Omega^{(1,2)}_1$ covering the arc  with the angle
$-\zeta<\psi<\pi+\zeta$, and $\Omega^{(1,2)}_2$ covering the arc
with the angle $\pi-\zeta<\psi<2\pi+\zeta$. Actually these maps
disappear  in the final formula for wave field \eqref {WF} because
both of them are nonsingular; they are used  like temporary
objects. From the other side changing these maps by their union
gives the tubular neighborhood  of $\Gamma_{t}$ which is not a
map.

3) Let us construct the covering of the front $\Gamma_{t}$
corresponding to the example of the axial symmetric case and for
time $t=t_{\rm{cr}}$, when the first focal point has just
appeared. Then we have only one focal point on
$\Gamma_{t_{\rm{cr}}}$, the one with the angle
$\psi^F(t_{\rm{cr}})=\pi/2$ (which follows from the symmetry of
the problem). Obviously we can choose covering consisting of two
maps: nonsingular $\Omega^{(1,2)}_1$ covering the arc  with the
angle $\pi/2 +2\zeta<\psi<2 \pi+\pi/2-2\zeta$ and singular
$\Omega^{(0,2)}_2$ covering the arc  with the angle $\pi/2
-\zeta<\psi<\pi/2+\zeta$.

4) Now we describe the covering of the front $\Gamma_{t}$ but for
$t>t_{\rm{cr}}$.  Denote by $\psi_{\pm}^F(t)$ the angles of the
left and right focal points. We can choose the maps in the
following way: the nonsingular  map $\Omega^{(1,2)}_1$, covering
the arc with the angle
$\frac{\pi}{4}-\zeta<\psi<\psi_{-}^F(t)-\zeta, \quad$ the singular
map $\Omega^{(0,2)}_2$, covering the arc with the angle
$\psi_{-}^F(t)-2\zeta<\psi<\psi_{-}^F(t)+2\zeta, \quad$ the
nonsingular map $\Omega^{(1,2))}_3$, covering the arc with the
angle $\psi_{-}^F(t)+\zeta<\psi<\psi_{+}^F(t)-\zeta \quad$ the
singular map $\Omega^{(0,2)}_4$, covering the arc with the angle
$\psi_{+}^F(t)-2\zeta<\psi<\psi_{+}^F(t)+2\zeta, \quad$ and the
nonsingular map $\Omega^{(1,2)}_5$, covering the arc with the
angle $\psi_{+}^F(t)+\zeta<\psi<\frac{3\pi}{2}+\zeta $.
\end{exam}
\begin{rem}
1) We mentioned that in the singular maps covering the  isolated
focal point with the angle $\psi^F(t)$ it is convenient to rotate
the coordinates in such a way that the new axis $x'_2$ has the
same direction as vector $\dot X(\psi^F(t))$. In the new
coordinate system the singular map always has a singular index
$I=(0,2)$ and focal (rotated) coordinates $(p'_1,x'_2)$. 2) One
needs to use the covering from item 3) of the previous example not
only for the critical time $t_{\rm{cr}}$ but also times for near
times $t$.
\end{rem}$\label{rotation}$.

\subsection{The Maslov index of a singular map.}

The last object we need is the Maslov index of chains of maps
$\{\Omega^{I_j}_j (t)\}$. To find  it one  has to fix some
nonsingular point $\mathbf{r}(\tilde \psi,\tilde \tau)$ in the
corresponding map $\Omega^{I_j}_j$ and construct their one of  the
following matrices $\mathbb{C}_\varepsilon
^{(1,0)},\mathbb{C}_\varepsilon ^{(0,2)}$ with the elements of
matrices $\mathcal{B}$, $\mathcal{C}$ defined in \eqref{BC}:
\begin{eqnarray*}\mathbb{C}_\varepsilon ^{(1,2)}=\mathcal{C}-i\varepsilon \mathcal{B}= \begin{pmatrix} \mathcal{C}
_{11 }-i\varepsilon B_{11} &\mathcal{C}_{12} -i\varepsilon
\mathcal{B}_{12}\\ \mathcal{C}_{21}-i\varepsilon
\mathcal{B}_{21} &\mathcal{C}_{22}-i\varepsilon
\mathcal{B}_{22}
\end{pmatrix}\\
\mathbb{C}_\varepsilon ^{(1,0)}=\\ \begin{pmatrix} \mathcal{C}
_{11 }-i\varepsilon B_{11} &\mathcal{C}_{12} -i\varepsilon
\mathcal{B}_{12}\\ (\mathcal{C}_{21}-i\varepsilon
\mathcal{B}_{21})\cos \eta+(\mathcal{B}_{21}+ i\varepsilon
\mathcal{C}_{21})\sin\eta &(\mathcal{C}_{22}-i\varepsilon
\mathcal{B}_{22})\cos \eta+(\mathcal{B}_{22}+i\varepsilon
\mathcal{C}_{22})\sin\eta
\end{pmatrix}\\ %&%#\
\mathbb{C}_\varepsilon ^{(0,2)}=\\
\begin{pmatrix}(\mathcal{C}_{11}-i\varepsilon
\mathcal{B}_{11})\cos \eta+(\mathcal{B}_{11}+ i\varepsilon
\mathcal{C}_{11})\sin\eta &(\mathcal{C}_{12}-i\varepsilon
\mathcal{B}_{12})\cos \eta+(\mathcal{B}_{12}+i\varepsilon
\mathcal{C}_{12})\sin\eta \\
\mathcal{C} _{21 }-i\varepsilon \mathcal{B}_{21} &\mathcal{C}_{22}
-i\varepsilon \mathcal{B}_{22}\end{pmatrix}\\
\mathbb{C}_\varepsilon ^{(0,0)}=
\\\begin{pmatrix}(\mathcal{C}_{11}-i\varepsilon
\mathcal{B}_{11})\cos \eta+(\mathcal{B}_{11}+ i\varepsilon
\mathcal{C}_{11})\sin\eta &(\mathcal{C}_{12}-i\varepsilon
\mathcal{B}_{12})\cos \eta+(\mathcal{B}_{12}+i\varepsilon
\mathcal{C}_{12})\sin\eta\\
(\mathcal{C}_{21}-i\varepsilon \mathcal{B}_{21})\cos
\eta+(\mathcal{B}_{21}+ i\varepsilon \mathcal{C}_{21})\sin\eta
&(\mathcal{C}_{22}-i\varepsilon \mathcal{B}_{22})\cos
\eta+(\mathcal{B}_{22}+i\varepsilon
\mathcal{C}_{22})\sin\eta\end{pmatrix} .
\end{eqnarray*}
These matrices are not generated for any $\eta\in [0,\pi/2]$ and
any positive $\varepsilon$ in the maps with index $(1,0)$, $(0,2)$
and $(0,0)$ respectively \footnote{As we mentioned above the
objects with singular index $(0,0)$ are not needed in the
considered problem we present $\mathbb{C}_\varepsilon ^{(0,0)}$
for completeness.}. Obviously $\mathbb{C}_\varepsilon
^{I}|_{\eta=0}=\mathbb{C}_\varepsilon
^{(1,2)}\equiv\mathcal{C}-i\varepsilon \mathcal{B}$
%{\bf the matrix $\mathbb{C}_\varepsilon ^{(1,2)}$ has not been introduced and the same matrix $\mathbb{C}_\varepsilon ^{I}|_{\eta=0}$ is used for these two equations so there should be some error}
and $\mathbb{C}_\varepsilon
^{I}|_{\eta=0}=\mathcal{C}^I-i\varepsilon \mathcal{B}^I$ and these
matrices determine a continuous non degenerate transition from the
matrix $\mathbb{C}_\varepsilon ^{(1,2)}$ to the matrix
$\mathcal{C}^I-i\varepsilon \mathcal{B}^{(1,2)}$. The
corresponding determinant $\mathbb{J}_{\varepsilon}^{(1,0)}=\det
\mathbb{C}_\varepsilon ^{(1,0)}$ or
$\mathbb{J}_{\varepsilon}^{(0,1)}=\det \mathbb{C}_\varepsilon
^{(0,2)}$ is not equal to zero. Let $m(\tilde \psi,\tilde \tau)$
be the Maslov index of the point $\mathbf{r}(\tilde \psi,\tilde
\tau)$. Then the index $\mathbf{m}(\Omega^{I_j}_j)$ of the map
$\Omega^{I_j}_j$ is
\begin{equation}\label{ind2}
\mathbf{m}(\Omega^{I_j}_j)=m(\tilde \psi,\tilde
\tau)+\frac{1}{\pi}\lim _{\varepsilon\to+0}
\rm{Arg}\mathbb{J}_{\varepsilon}^{I}|^{\eta=\pi/2}_{\eta=0}.
\end{equation}
This definition {\it does not depend $\mod 4$ on the choice of the
point $(\tilde \psi,\tilde \tau)$ in a given map}. The calculation
of the index $\mathbf{m}(\Omega^{I_j}_j)$ could be technically
complicated even in quite simple situations. But taking into
account the fact that we can restrict ourselves to maps with
$I=(1,0)$ and  $I=(0,2)$  it is possible to simplify the
application of this formula.
\begin{lemma}
One can always find  a nonsingular point $\mathbf{r}(\tilde
\psi,\tilde \tau)$ in the map $\mathbf{m}(\Omega^{I_j}_j)$ with
$I_j=(1,0)$ or  $(0,2)$ such that the sign Jacobian $J(\tilde
\psi,\tilde \tau)$ coincides  with the sign of the Jacobian
$J^{I_j}(\psi,\tau)$ in this map. Then the second term in
\eqref{ind2} is equal to zero and
$$\mathbf{m}(\Omega^{I_j}_j)=m(\tilde \psi,\tilde \tau).$$
As  the sign of the Jacobian  $J^{I_j}$ does not depend on a point
in the $\Omega^{I_j}_j$, thus one can find it in any point, for
instance in a focal one.
\end{lemma}

\textsc{Proof}. It is obvious that the second term in \eqref{ind2}
can be equal to $0,1$ or $-1$ only. Consider for $\varepsilon$ the
Jacobian $\mathbb{J}_{0}^{I}$. A simple calculation gives
$\mathbb{J}_{0}^{I}= J\cos \eta+J^{I}\sin \eta$. In the interval
$[0,\pi/2]$ this function has no zero if $J$ and $J^{I}$ have  the
same signs and one zero in the opposite situation. Hence including
the parameter $\varepsilon$ gives only the  rule of bypassing %bypass%New bypass Endnew OBHODA {\bf circle rule? the zero point of what? it is not clear} 
of the zero
point on the complex plane, and it is not necessary to use this
rule in the case when $JJ^{I}>0$ Thus one obtains
$\lim_{\varepsilon\to+0}
\rm{Arg}\mathbb{J}_{\varepsilon}^{I}|^{\eta=\pi/2}_{\eta=0}=0$ if
a point $\mathbf{r}(\tilde \psi,\tilde \tau)$ is chosen in the way
prescribed in the Lemma. From the other side, according to Lemma 3
and its Corollary, the existence of the focal point in any focal
map means the existence of nonsingular points with positive and
negative signs of the Jacobian $J$. $\Box$
\begin{cor} The index $\mathbf{m}(\Omega^{I_j}_j(t))$ of the singular
map  $\Omega^{I_j}_j$ coincides with the index of any nonsingular
point $m(\psi,t)$ on the  front $\Gamma_t$ where the Jacobians $J$
and $J^{I_j}$ have the same sign.
\end{cor}
Let us illustrate this lemma by the calculations of indices of
maps from the examples considered above.

\begin{exam}

1) Consider the maps from the example \eqref{Gamma0}. Let us take
the points on the maps $\Omega^{I_j}_j, j=1,2,3,4$ with
coordinates $\psi, \tau$, $\tau>0$. We put $m(\psi,\tau)=0$.
According to Corollary from  Lemma 4 % {\bf I am not sure about this lemma}
$J(\psi,\tau)>0$.
From this and \eqref{jac2} we find
$\mathbf{m}(\Omega^{I_j}_j)$=0 for each $j=1,2,3,4$.
\end{exam}

\begin{exam}

 2)Consider the maps from 3), 4) of example \eqref{Gamma0}. Before
let us come back to formulas \eqref{3.4a}. As we mentioned in
remark \eqref{rotation} it is convenient  to rotate the
coordinates. Then in the new coordinates in the focal point $X^F$
with coordinates $(\psi^F(t),t)$
$$\dot X'=P' C^2(X^F)/C_0
\begin{pmatrix}0 \\q a n \end{pmatrix},\quad  \text{and} \quad
P'_{\psi}=\begin{pmatrix}-q b(n+1)\\0 \end{pmatrix}.
$$
Hence $J'^{(0,2)}=abC^2(X^F)q^2 \psi n(n+1)/C_0$ and the sign of
$J'^{(0,2)}$ coincides with the sign of the product $ab$. For both
examples 3.) 4.) the sign is negative (see Fig. ) %{\bf check the references inserted by me, also I am not sure where one can see that this sign is negative }, 
thus the index of the map
$\Omega^{(0,2)}_2$ in 3 of example \eqref{Gamma0}, and indices of
the maps $\Omega^{(0,2)}_2$, $\Omega^{(0,2)}_4$ in 3 of example
\eqref{Gamma0} are $-1$.
\end{exam}

%ENDNEW

\subsection{ Germs of Lagrangian manifolds and their properties.}
The geometrical objects described above  one can meet in many asymptotical problems having fast oscillation solution. We see that the solution of considered problem decays quit rapidly outside of some neighborhood of the front $\gamma_t$. This gives the opportunity to use the ideas of boundary layer
 expansions \cite{VishikLust}-\cite{MasOm} and  the 
 ``complex germ theory'' \cite{M2}. Their geometrical realization contains   in a change Lagrangian band $M^2_t$ by its  linearization (germ) near the curve $\Gamma_t$.
%First let us  define the germ of Lagrangian manifold.
\begin{definition} For each fix $t$ we call the  linear germ corresponding
to the manifold $M^2_T$ a vector fiber bundle in the phase space
with base coinciding with the front $\Gamma_t=(x=X(\psi,t),
p=P(\psi,t))$ and fibers generated by the vectors $\dot X,\dot P$.
\end{definition}
Denote $\alpha\in \mathbb{R}$ the coordinate on the bundle (which is a linear  analog to a proper time $\tau$), then
we can define a family of  manifolds $\Lambda^2_t$ as a strip in a
neighborhood of the front $\Gamma_t$ in the phase space
$\mathbb{R}^4_{p,x}$
\begin{equation}
\Lambda^2_t=\{p={\mathrm{P}}(\psi,t,\alpha)\equiv P(\psi,t)+\dot
P(\psi,t)\alpha,\quad x={\mathrm{X}}(\psi,t,\alpha)\equiv X(\psi,t)+\dot
X(\psi,t)\alpha\},\label{germ}
\end{equation}
where $\psi \in S^1=[0,2\pi],\quad |\alpha|<\alpha_0$ are the
coordinates in $\Lambda_t^2$. It is easy to see that $M^2_t$ can be
approximated by $\Lambda^2_t$, and that the parameter $\alpha$ is
used for linearizing the functions
$X(\psi,t+\alpha),P(\psi,t+\alpha)$ and that it defines a shift of
time near the front $\Gamma_t$. Taking into account this fact it
is easy to prove the following proposition.
\begin{lemma} 1) With an error of the order $O(\alpha ^2)$ the
manifold $\Lambda_t^2$ is obtained from $\Lambda_0^2$ by means of
a shift of time $t$ along the trajectories of the phase flow with
the Hamiltonian $ \mathcal{H} = H|p|C(x)$. \label{lem:1} 2)  For
the matrices ${\mathrm{B}}(\psi,t,\alpha)=\frac{\pa
\mathrm{P}}{\pa(\alpha,\psi)},\,\,$ ${\mathrm{C}}(\psi,t,\alpha)=\frac{\pa
\mathrm{X}}{\pa(\alpha,\psi)}$ the following equalities are true
$\mathrm{B}=\mathcal{B}+O(\alpha),\mathrm{C}=\mathcal{C}+O(\alpha)$; ${}^t
\mathrm{C}\mathrm{B}={}^t \mathrm{B}\mathrm{C}+O(\alpha)$,
where as before $\mathcal{B}=\frac{\pa {P}}{\pa(t,\psi)},\quad
\mathcal{C}=\frac{\pa {X}}{\pa(t,\psi)}$. The last equality
means that the manifold (band) $\Lambda^2_t$ is (almost)
Lagrangian mod $O(\alpha)$, the statement 1) means that it is
(almost) invariant mod $O(\alpha)$.
\end{lemma}

\textsc{Proof}. Consider the Hamilton equations
$\dot x=  \mathcal{H}_p, \dot p= - \mathcal{H}_x$.
We expand the derivatives of the Hamiltonian $ \mathcal{H}$ around the
point $X(t,\psi)$, $P(t,\psi)$
\begin{gather*}
 \mathcal{H}_p=  \mathcal{H}_p(X(\psi,t), P(\psi,t))+  \mathcal{H}_{pp}\alpha\dot P +\mathcal{H}_{px}\alpha \dot X +O(\alpha^2)\\
 \mathcal{H}_x=  \mathcal{H}_x(X(\psi,t), P(\psi,t))+ \mathcal{H}_{xp}\alpha\dot P+  \mathcal{H}_{xx} \alpha\dot X +O(\alpha^2)
\end{gather*}
and substitute in the Hamilton equations and we get the result.
The second proposition is obvious and use the variational system
for the evolution of $(x,p)$:
\begin{gather*}
 \dot X + \alpha \ddot X- (\mathcal{H}_p(X(\psi,t), P(\psi,t))+  \mathcal{H}_{pp}\alpha\dot P +\mathcal{H}_{px}\alpha \dot X )=O(\alpha^2)\\
 \dot P + \alpha \ddot P+ \mathcal{H}_x(X(\psi,t), P(\psi,t))+ \mathcal{H}_{xp}\alpha\dot P+  \mathcal{H}_{xx} \alpha\dot X= O(\alpha^2)
\end{gather*}
and we get the result. $\Box$

As the germ $\Lambda^2_t$ is some approximation of $M^2_t$ almost all previous proposition, geometrical definition and  construction (like Maslov and Morse index) related to the band $M^2_t$ is true for the band (germ) $\Lambda^2_t$. From the other side obviously one does not need any additional objects besides the family of curves (fronts in the phase space) $\Gamma_t$ to construct both the Lagrangian bands $M^2_t$ and their germs $\Lambda^2_t$. It also follows from  formulas \eqref{WF},\eqref{5.2},\eqref{etafoc} that leading term of the solution $\eta$ based also only on these objects. Nevertheless the proof of \eqref{WF},\eqref{5.2},\eqref{etafoc} needs something more, and it seems that sometimes technically instead of the germ $\Lambda^2_t$ it is convenient to consider also the other germ
(fiberbundle) namely
\begin{equation}
\widetilde {\Lambda}^2_t=\{p=\widetilde{\mathrm{P}}(\psi,t,\alpha)\equiv P(\psi,t)+\big(\dot
P(\psi,t)- \lambda(\psi) P\big)\alpha,\quad x={\mathrm{X}}(\psi,t,\alpha)\equiv X(\psi,t)+\dot
X(\psi,t)\alpha\}\label{germ1}
\end{equation}
where (see Lemma 2) $\lambda=\langle \frac{\pa C}{\pa x}(0),\mathbf{n}(\psi)\rangle$.
This germ also implies the  matrices $\widetilde{\mathrm{B}}(\psi,t,\alpha)=\frac{\pa
\mathrm{P}}{\pa(\alpha,\psi)},\,\,$ $\widetilde{\mathrm{C}}(\psi,t,\alpha)={\mathrm{C}}(\psi,t,\alpha)=\frac{\pa
\mathrm{X}}{\pa(\alpha,\psi)}$. But now $\widetilde{\mathrm{B}}=
\widetilde{\mathcal{B}}+O(\alpha)$, where ${\mathcal{B}}=(\dot P-\lambda P,P_{\psi})$.
After analysis of our previous consideration it is possible to prove the following statement.
\begin{lemma} All previous proposition concerning the germ $\Lambda^2_t$ and
matrices $\mathcal{B},\mathcal{C}$ are true for the germ $\widetilde {\Lambda^2_t}$ and matrices $\widetilde{\mathcal{B}},\mathcal{C}$.
\end{lemma}

%{\bf I AM NOW CORRECTING THE NEXT (LAST) SECTION, I TRIED to INTRODUCE ALL CORRECTION of BRUNELLO and SERIOZHA Sekerzh-Z.}

\section{Geometrical asymptotic solution and the Maslov canonical operator}.
 The central mathematical result of our paper is the observation that the
asymptotic solution of the problem \eqref{WEq} can be represented
as an integral over $d\rho$ of the canonical Maslov operator with
``semiclassical'' parameter $h=l/\rho$, defined on the appropriate
family of Lagrangian manifolds $\Lambda^2_t$, and acting on the
function $V$ \eqref{Sour} defining the initial localized
perturbation. In some neighborhood of the front line $\gamma_t$
the final formula has the form: %{\bf NEW Formula}
\begin{equation}
\eta=\rm{Re}\big(\sqrt{\frac{\mu C_0}{2\pi i}}
\int_0^{\infty}K_{\Lambda^2_t}^{\mu/ \rho}\big(\sqrt{\rho} \tilde V (\rho \mathbf{n}(\psi)\big) d\rho+
o(\mu), \label{5.3}\end{equation}
and $\eta=o(\mu)$
outside of this neighborhood. The initial data \eqref{Sour}, the
representation of the asymptotic solution \eqref{WF},\eqref{5.2}
out of the neighborhood of the focal points as well as the future
representation of the solution in the neighborhood of the focal
points is only a realization of \eqref{5.3} in the corresponding
domain of $\mathbb{R}^2_x$.  As we said before the integral over parameter $\rho$ in
\eqref{5.3} plays a very important role: it implies the decay of
the function $\eta$ outside a neighborhood  of the front and in
turn allows one to simplify the objects and  formulas appearing in
the construction of the Maslov canonical operator. This
simplification is based on the mentioned ideas of the ``complex germ
theory'' \cite{MaslovOperMethods,M2}, but in a simpler 
``boundary layer'' version. 
As we said before from the geometrical point of 
view it means that we can use germs $\Lambda^2_t$  
or $\widetilde\Lambda^2_t$ instead of the Lagrangian band 
 $M^2_t$  in  \eqref{5.3}. In next subsections 
 we shall describe the functions and other objects 
 determining the operator $K_{\Lambda^2_t}^{\mu/ \rho}$.

\subsection{The functions on Lagrangian bands $M^2_t$.}

{\bf a.The  action-function.} The Lagrangian  property  allows one to define on the family  $M^2_t$,
function $s(\psi,t,\alpha)$ satisfying the equation $ds=\langle
P,dX\rangle\big|_{M^2_t}$.
\begin{lemma} The phases $(\psi,\tau)$ on $M^2_t$ is equal to $C_0\alpha$.
\end{lemma}
\textsc{Proof}. First let us find $s$ on $M^2_0$. Then the  coordinate $\alpha$ is the proper time: $\tau=\alpha$. But in this case  we have
$$\int_{(0,0)}^{(\psi,\tau)}\langle P,dX\rangle=
\int_{(0,0)}^{(0,\tau)}\langle P,\dot X\rangle
dt+\int_{(0,\tau)}^{(\psi,\tau)}\langle P, X_\psi\rangle d\psi.$$ The
second term in the last expression is equal to zero according to
Lemma 2. Changing $\dot X$ by the right hand side from system
\eqref{Ham3} and using the integral of motion \eqref{Ham4}, we
find (using the proper time $\tau$)
$s(\psi,\tau)=C_0\tau=C_0\alpha.$ According to 
\cite{MaslovAsymptMethods,MaslovFedoryuk,DobrZh}
to construct the action on the band $M^2_t$ one has to add $\int_0^t\mathcal{L}dt$ to $C_0\alpha$, where the Lagrangian
$\mathcal{L}=\langle p,\mathcal{H}\rangle-\mathcal{H}$. But for the wave equation  $\mathcal{L}=0$,
which gives the first proposition of  Lemma.$\Box$

The phase $s$ implies the phases associated with projection to singular  maps $\Omega^I_j$:
\begin{equation*}%\label{phase1}
s^{(1,0)}(\psi,\tau)=C_0\tau-P_2 (\psi,\tau)X_2(\psi,\tau),
s^{(0,2)}(\psi,\tau)=C_0\tau-P_1
(\psi,\tau)X_1(\psi,\tau).\end{equation*}

The initial source function $V(y)$ implies on $M^2_t$ the function (more precisely the family of function depending on parameter $\rho\in (0,\infty)$)
\begin{equation}\label{f}
f(\rho,\psi)=\sqrt{\rho} \tilde V (\rho \mathbf{n}(\psi).
\end{equation}

We need also the smooth  cut off function $\mathbf{e}_0(\alpha)$,
$\mathbf{e}_0(\alpha)=1$ for $|\alpha|<\alpha_0$, and
$\mathbf{e}_0(\alpha)=0$ for $|\alpha|>2\alpha_0$ where $\alpha_0$ is
some small enough positive number. The product $f(\rho,\psi)\mathbf{e}_0(\alpha)$ gives the finite function on the band $M^2_0$. We continue it on all family $M^2_t$ assuming that it doesnot depend on time $t$.

{\bf b. Functions %$(\psi^I_j(x^I),\tau ^I_j(x^I))$, action- and source functions  and Jacobians
in the maps $\Omega^I_j$.}
In each map $\Omega^I_j$ the Jacobians $J^I$ are not equal to zero. This means that one can construct the smooth solutions
$(\psi^I_j(x^I),\tau ^I_j(x^I))\in \Omega ^{I_j}$
of the system of equations
\begin{equation}\label{xp}
    X^I(\psi,\tau)=x^I.
\end{equation}
Let us emphasize that it could exists several angles $\psi^I_j(x^I)$, corresponding to one vector $x^I$.
Although the manifold $M^2_t$ is invariant with
respect to the shift $g^t_{\mathcal{H}}$ it does not mean that
there is no dynamics on $M^2_t$. It only means that turning on the
time dependence we transform the objects related to $M^2_t$ in a
special way. Namely let us use from the beginning  the coordinate
$\alpha$ instead of the proper time $\tau$. We have already
mentioned that the points $\mathbf{r}(\psi,\alpha)\in M^2_T$ after
the action of the transform $g^t_{\mathcal{H}}$ passes to the
points $\mathbf{r}(\psi,\alpha)\in M^2_t$ with shifted coordinate
$\tau=\alpha+t$, but with the same angle $\psi$. Thus the
equations \eqref{xp} are changed by the equations
\begin{equation}\label{xp1}
    X^I(\psi,\alpha+t)=x^I;
\end{equation}
The following trivial proposition is  very useful.
\begin{lemma}

Let $\psi_j(x^I),\tau_j(x^I) $ be the solution of the equations
\eqref{xp} in the map $\Omega_j^{I_j}(t)$ and let the point
$\mathbf{r}(\psi,\alpha+t)\in M^2_t$ with coordinates
$\psi,\alpha+t$ belong to the same map $\Omega_j^{I_j}$. Then the
angle component $\psi_j$ of the solution of the equation
\eqref{xp1} does not depend on $t$: $\psi_j=\psi_j(x^{I_j})$ and
$\alpha_j=\alpha_j(x^{I_j},t)$  is
\begin{equation}\label{xp3}
 \psi_j(x^{I_j},t)=\psi_j(x^{I_j}),\quad \alpha_j(x^{I_j},t)=\tau_j(x^{I_j})-t,
\end{equation}
\end{lemma}
\textsc{Proof} is obvious $\Box$.

Using \eqref{xp1} and \eqref{xp3} we can rewrite  action-functions and  the Jacobians in the  coordinates $x^I$.
The behavior of the functions on $M^2_t$ is different with respect
to the shift $g^t_{\mathcal{H}}$. Namely the function
$s=\alpha$, and  the functions $f$,$\mathbf{e}$  are constant ,
this means that for each $t$ in the coordinates $\psi,\alpha$ one
has the same form. On the contrary the functions $s^{I}$ and all
the Jacobians \footnote{To simplify  notation we do not introduce
a new symbol for time-shifted Jacobian.}
 $J^I$ take the forms:
\begin{gather}\nonumber s^{(1,0)}(\psi,\alpha,t)=C_0\alpha-P_2 (\psi,\alpha+t)X_2(\psi,\alpha+t),\\
s^{(0,2)}(\psi,\alpha+t)=C_0 \alpha-P_1
(\psi,\alpha+t)X_1(\psi,\alpha+t),\label{phase2}\end{gather}
\begin{equation}\label{jac3}J^I=J^I(\psi,\alpha+t).\end{equation}
Now in  the maps $\{\Omega^{I_j}_j(t)\}$ we want to pass from
coordinates $\psi,\alpha$ to coordinates $x^{I_j}$. This gives us
the  actions, Jacobians etc. in the coordinates $x^{I_j}$:
\begin{gather}\nonumber S^{(1,2)}_j(x_1,x_2)=\tau_j(x_1,x_2)-C_0t,\\
\nonumber S^{(1,0)}_j(x_1,p_2)=\tau_j(x_1,p_2)-
p_2 X_2(\psi_j(x_1,p_2),\tau_j(x_1,p_2))-C_0t,\\
\nonumber S^{(0,2)}_j(p_1,x_2)=\tau_j(p_1,x_2)-p_1
X_1(\psi_j(p_1,x_2),\tau_j(p_1,x_2))-C_0t,\\ \nonumber J^{I_j}_j(x^{I_j})=
J^{I_j}(\psi_j(x^{I_j}),\tau_j(x^{I_j})),\\
\mathbf{e}^t=\mathbf{e}(\tau_j(x^{I_j})-t).\label{jac4}\end{gather}
Let us emphasize that the complicated notations only reflect the
situation: each map has it own number $j$ and index of singularity
$I_j$. (See examples below).

Finally we need to introduce the partition of unity with the maps $\Omega^{I_j}_j(t)$ covering $\Gamma_t$:  the set of  smooth functions
$\mathbf{e}_j(\psi,t)$ associated with the covering
$\{\Omega^{I_j}_j(t)\}$: $\rm{supp}\mathbf{e}_j(\psi)\in
\Omega^{I_j}_j(t)$, $\sum_j\mathbf{e}_j(\psi)=1$.

%e) the partial inverse $h$-Fourier transforms $$[\mathcal{F}^{-h}_{p_1\to x_1}\chi(p_1,x_2)](x_1,x_2)= \frac{i}{\sqrt{2\pi}} \int_{-\infty}^{-\infty}\chi(p_1,x_2)e^{\frac{ip_1x_1}{h}}dp_1,$$ and $$[\mathcal{F}^{-h}_{p_2\to x_2}\chi(x_1,p_2)](x_1,x_2)=\frac{i}{\sqrt{2\pi}} \int_{-\infty}^{-\infty}\chi(x_1,p_2)e^{\frac{ip_2x_2}{h}}dp_2.$$

\subsection{The time-dependent canonical Maslov operator on the
invariant manifold $M^2_t$.}$\label{invariant.}$

Now everything is ready to determine the canonical Maslov operator
$K_{M^2_t}^h$, acting on the function $f(\rho,\psi)
\mathbf{e}(\alpha)$, which is constant one on the trajectories of
system \eqref{Ham3} and depending on the parameter $h>0$. It means
that this function is the same in all points
$P(\psi,t+\alpha),X(\psi,t+\alpha)$. Let $\{\Omega^{(I_j)}_j\}$ be
a covering of the curve $\Gamma_t$. Let us divide the set of
indices $\{j\}$ into three parts
$\{\mathbf{j}(1,2),\mathbf{j}(1,0),\mathbf{j}(0,2)\}$
corresponding to the maps with indices of singularity
$(1,2)$,$(1,0)$ and $(0,2)$ respectively. We put
\begin{gather} \nonumber \Psi(\rho,x_1,x_2,t)= K_{M^2_t}^h(f\mathbf{e})\equiv\\
\nonumber \sum_{j\in
\mathbf{j}(1,2)}e^{-\frac{i\pi}{2} m(\psi_j(x_1,x_2))}
\frac{\exp{\frac{iS_{j}^{(1,2)}(x_1,x_2,t)}{h}}}{\sqrt{|J_j^{(1,2)}(x_1,x_2)|}}
f(\rho,\psi)
\mathbf{e}_{j}(\psi)|_{\psi=\psi_{j}(x_1,x_2)}\mathbf{e}(\tau_{j}(x_1,x_2,t)-t)
+\\ \nonumber \sum_{j\in
\mathbf{j}(1,0)}e^{-\frac{i\pi}{2}\mathbf{m}(\Omega_j^{(1,0)})}
\frac{i}{\sqrt{2\pi h}}
\int_{-\infty}^{+\infty}\frac{\exp{\frac{i(S_{j}^{(1,0)}(x_1,p_2,t)+x_2p_2)}{h}}}
{\sqrt{|J_j^{(1,0)}(x_1,p_2)|}}
f(\rho,\psi)\mathbf{e}_{j}(\psi)|_{\psi=\psi_{j}(x_1,p_2)}
\mathbf{e}(\tau_{j}(x_1,p_2,t)-t)
dp_2+\\
\sum_{j\in
\mathbf{j}(0,2)}e^{-\frac{i\pi}{2}\mathbf{m}(\Omega_j^{(0,2)})}
\frac{i}{\sqrt{2\pi h}}
\int_{-\infty}^{+\infty}\frac{\exp{\frac{i(S_{j}^{(0,2)}(p_1,x_2,t)+x_1p_1)}{h}}}
{\sqrt{|J_j^{(0,2)}(p_1,x_2)|}}
f(\rho,\psi)\mathbf{e}_{j}(\psi)|_{\psi=\psi_{j}(p_1,x_2)}
\mathbf{e}(\tau_{j}(p_1,x_2,t)-t) dp_1 \label{KO1}
\end{gather}
where $S^I_j$ and $J^I_j$ are defined in \eqref{jac4}.
 Now we can construct the asymptotic solution $\eta$ to the problem \eqref{WEq}.
 We put in the last formula $h=\rho/l$.

\begin{theorem}
1) For any $T$ independent of  $\,\mu=l/L$, the solution $\eta$ to the
problem \eqref{WEq} in the interval $t\in [0,T]$ has the form:
\begin{equation}\label{sol1}
    \eta=\eta_{\rm{as}}+o(\mu),\quad
   \eta_{\rm{as}}= \sqrt{\frac{\mu C_0}{2 \pi }}
   \rm{Re}\big(e^{\frac{-i \pi}{4}}\int_0^\infty\Psi(\rho,x_1,x_2)\big)d\rho.
\end{equation}
This  asymptotic, apart from terms of the order  $O(\mu)$,  does
not depend on the choice of the covering $\{\Omega_j^{I_j} \}$,
and functions $\mathbf{e_j^t}$, $\mathbf{e}$.

2) For each time $t\in [O,T]$  the function $\eta$ is localized in
a neighborhood of the front: the function $\eta$ is equal $O(\mu)$
outside some neighborhood of the front $\gamma_t.$
\end{theorem}

\textsc{Sketch of half of Proof.} Using the results 
\cite{MaslovAsymptMethods,MaslovFedoryuk,
MaslovOperMethods,DobrZh,DobrMZhSh} one can show that the function \eqref{KO1} is a leading term of some asymptotic solution $\Psi^k$ $\rm{mod}O(h^k)$ to original equation \eqref{WEq}, where $k$ is an arbitrary big by fixed integer number. We introduce the smooth cut off function $\mathbf{g}(y)$: $\mathbf{g}(y)=0$ for  $y\leq 1/2$ and $\mathbf{g}(y)=1$ for
$y\geq 1$. Multiplying $\Psi^k$ by $\mathbf{g}(\rho/\mu^{1/2})$ and integrating the product by $d \rho$ we obtain that the result is asymptotic solution of \eqref{WEq} $\rm{mod}O(\mu^2)$. Then as in 
 \cite{DobrMZhSh} we show that the influence of the term   $\int_0^1(1-\mathbf{g}(\rho/\mu^{1/2}))\Psi^k|_{t=0}d\rho$   into the
solution \eqref{WEq} is $o(\mu)$, and hence the function $\eta_{as}$ from \label{sol1} is a leading term of some asymptotic solution of \eqref{WEq}. Now we need to check the conditions \eqref{WEq1}. But it is better to do after simplification of function \label{sol1} and we shall do it in the next subsection.

\subsection{The germ  $\Lambda_t^2$ of the manifold $M^2_t$ and the
simplification of the asymptotic.}$\label{germ}$

Since the function \eqref{sol1} decays quite rapidly when the
point $x$ goes away from the front, it is possible to change the
functions $S^{I_j},|J_j^{I_j} $ in a neighborhood of the front by
their Taylor expansions. The nice fact is that one does not change
the accuracy $O(\mu)$ in formula \eqref{sol1} using only the zero,
first terms and sometimes second terms of the Taylor expansions of
the phases, and zero terms in the other functions. All these
expansions are expressed via the vector functions
$(P(\psi,t),X(\psi,t))$  and matrices
$\mathcal{B}(\psi,t),\mathcal{C}(\psi,t)$.

We need  the Taylor expansion in the following form. Let the
equations $(y_1,y_2)=(Y_1(\psi),Y_1(\psi))$ determine  a smooth
curve $\Upsilon$ in some domain in  $R^2_y$ and $\Phi(y_1,y_2)$ be
a smooth  function in some neighborhood $\mathcal{D}$ of
$\Upsilon$. Let $q(\psi)$ be the smooth family of nonzero vectors
with components $(q_1,q_2)$  transversal to the curve $\Upsilon$.
%{\bf why not choosing $q(\psi)$ orthogonal to the curve?} {\bf Sergey's comment: because sometimes it gives simpler representation of the solution}
This means that the vectors $q(\psi)$ and $Y_\psi(\psi)$ are not
parallel. The parameter $\psi$ on $\Upsilon$ and the family of
vectors $q(\psi)$ define the curvilinear system in some
neighborhood of $\Upsilon$: each point $y$ in this neighborhood
can be characterized by two values: $\psi(y)$ and the length (with
the sign) $
z=\langle y-Y(\psi(y)),q(\psi(y))\rangle/(q(\psi(y)))^2$ of the vector
$y-Y(\psi(y))$. To find the value of $\psi(y)$ one has to solve
the equation \begin{equation}\label{y1} \langle
y-Y(\psi),q_\perp(\psi)\rangle\equiv
(y_1-Y_1(\psi))q_2(\psi)-(y_2-Y_2(\psi))q_1(\psi)=0.
\end{equation}

%{\bf I can understand this equation if the vector $q(\psi)$ is parallel to the surface $\Upsilon$ } {\bf Sergey Comments.  The vector $q$ is transversal to the curve, and $x-X$ is parallel to $q$. From the picture you can see the last equation.}

\begin{lemma} The following expansion is valid:
\begin{equation}\label{exp}
    \Phi(y)=\Phi(Y)+
\langle\frac{\pa\Phi}{\pa y}(Y),(y-Y)\rangle+\frac{1}{2}
\langle(y-Y),\frac{\pa^2\Phi}{\pa
y^2}(Y)(y-Y)\rangle|_{Y=Y(\psi(y))}+O((y-Y(\psi(y)))^3)
\end{equation}
\end{lemma}
\textsc{Proof} follows from the 1-D Taylor expansion of the
function $\Phi(Y+q z)$ with respect to variable $z$ $\Box$.

Now we want to apply this lemma to the phases $S^{I_j}_j$ and
Jacobians $J^{I_j}_j$  in \eqref{jac4} and \eqref{sol1}. The variable $y$ are
$x^I$, the curve $\Upsilon=\{x^I=X^I(\psi,t)\}$, thus the solution
$\psi$ will depend also on time t. We need the first and second
derivatives of $S^{I}$ in the points $x^I=X^I(\psi,t)$. From the
general theory of Hamilton-Jacobi equation
\cite{Arnold,MaslovTeorVoz,MaslovFed} it follows
$$\frac{\pa S^{I}}{\pa x^I}=P^I(\psi,t),\quad\frac{\pa^2 S^{I}}{\pa (x^I)^2}=
\frac{\pa P^I}{\pa x^I}(\psi,t)\equiv \mathcal{B}^I (\psi,t) (\mathcal{C}^I(\psi,t))^{-1},$$
where %$\psi(x^I),\tau(x^I)=t+\alpha(x^I)$ is a solution to system \eqref{},and
the matrices $\mathcal{B}^I,\mathcal{C}^I$ are defined in
\eqref{matr}. Now let us choose the  vector $q$ as following. In
the case $I=(1,2)\,\,$ $q=\dot X(\psi,t)=(X_\psi)_\perp$, then
equation \eqref{y1} is equation. \eqref{xp1}. In the case
$I=(0,2)\,\,$ $q={}^t(0,1)$, then equation \eqref{y1} is
\begin{equation}\label{psi1}
    P_1(\psi,t)=p_1.
\end{equation}
In the case $I=(1,0)$\, $q={}^t(1,0)$, then Eq.\eqref{y1}
is
\begin{equation}\label{psi2}
    P_2(\psi,t)=p_2.
\end{equation}
 We denote $\psi^j(x^{I_j},t)$ the solution of these equations in the
map with the number $j$. \footnote{These solutions are different
from the solutions $\psi_j(x^{I_j})$ of equation \eqref{xp3}, we
use almost the same symbol to simplify the notation.} Then after
some algebra we obtain the following formulas in the maps with
numbers $j$
\begin{gather}\nonumber S^{I_j}_j(x^I,t)=\{\mathcal{S}^{I_j}_j(\psi,t)+O(x^{I_j}-
X^{I_j}(\psi,t))^3\}|_{\psi=\psi^j(x,t)},\\
J^{I_j}_j(x^{I_j},t)=\{\mathcal{J}^{I_j}_j(\psi,t)+O(x^{I_j}-
X^{I_j}(\psi,t))\}|_{\psi=\psi^j(x,t)},
\label{sglob} \end{gather} here %$\psi=\psi^j(x^{I_j},t)$ and
\begin{gather}\nonumber
\mathcal{S}^{(1,2)}_j(\psi,t)=\langle P(\psi,t),(x-X(\psi,t)\rangle-
\frac{1}{2}\langle P(\psi,t)),C_x(X(\psi,t))\rangle (x-X(\psi,t)^2%]|_{\psi=\psi^j(x,t)}
,\\
\mathcal{J}^{(1,2)}_j=\det (\dot
X,X_{\psi})(\psi,t)%|_{\psi=\psi^j(x,t)}
\label{s12}
\\ \nonumber
\mathcal{S}^{(0,2)}_j(\psi,t)= -P_1(\psi,t)X_1(\psi,t)+
P_2(\psi,t)(x_2-X_2(\psi,t))+\\ \nonumber \frac{1}{2}
(x_2-X_2(\psi,t))^2\frac{\dot P_1 P_{2\psi}-\dot P_2P_{1\psi}  }
{\dot P_1  X_{2\psi}-\dot X_2 P_{1\psi}}%]|_{\psi=\psi_j(p_1,t)}
\\
\mathcal{J}^{(0,2)}(\psi,t)=\det
\mathcal{C}^{(0,2)}(\psi,t)%|_{\psi=\psi^j(p_1,t)}
=
(\dot P_1 X_{2\psi}-\dot X_2 P_{1\psi})(\psi,t)\label{s02}\\
\nonumber \mathcal{S}^{(1,0)}(\psi,t)=-P_2(\psi,t)X_2(\psi,t)+
P_1(\psi,t)(x_1-X_1(\psi,t))+\\ \nonumber \frac{1}{2}
(x_1-X_1(\psi,t))^2\frac{\dot P_1 P_{2\psi}-\dot P_2P_{1\psi}}
{\dot X_1  P_{2\psi}-\dot P_2 X_{1\psi}}%|_{\psi=\psi^j(p_2,t)}
\\\mathcal{J}^{(1,0)}(\psi,t)=\det
\mathcal{C}^{(1,0)}(\psi,t)%]|_{\psi=\psi^j(p_2,t)}
= ({\dot X_1
P_{2\psi}-\dot P_2 X_{1\psi}})(\psi%^j(p_2
,t),t)\label{s01}
\end{gather}
\begin{rem} It is important that the last formulas do not depend on
the choice of the vector $q$ with the same accuracy they are valid.
\end{rem}
\begin{theorem}
The proposition of Theorem 1 is valid if one changes in the
formulas \eqref{KO1},\eqref{sol1} $S^{I_j}_j$  by
$\mathcal{S}^{I_j}_j$, $J^{I_j}_j $, by $\mathcal{J}^{I_j}_j$,
$\psi_{j}(x^{I_j})$ by $\psi^{j}(x^{I_j},t)$,
$\mathbf{e}(\tau_{j}(x^{I_j},t)-t)$ by
$\mathbf{e}( |x^{I_j}-X^{I_j}(\psi^{j}(x^{I_j},t)|)$. 2)
In the singular maps in formulas \eqref{KO1} one can change the
integration over  $p_j\in (-\infty,\infty)$ by the integration
over {\it the angle} $\psi\in \Omega_j^{I_j}$, putting
$p_j=P_j(\psi,t)$ and $dp_j= \frac{\pa P_{j}}{\pa
\psi}(\psi,t)d\psi$ adjusting the limits in the integral with these change.
\end{theorem}
\textsc{The idea of Proof.} The proof in regular maps is no more but the
Taylor expansion of regular components in \eqref{KO1}, \eqref{sol1}
with respect to distance from $\gamma_t$. The proof in the focal maps based on Taylor expansions but also on  estimates of some rapidly oscillating integrals.

\subsection {Derivation of formulas from Theorems 1-3}
Now let us apply this Theorem in different cases.

{\bf a. A verification of initial data}
 We shall start
from $t=0$. It was shown in Example \eqref{Gamma0} that one can
cover the neighborhood of the curve $\Gamma_0$ by maps
$\Omega^{(1,0)}_{1,3}$ and $\Omega^{(0,2)}_{2,4}$. Taking into account
the definition of the functions $\mathbf{e}_{j}$ the equalities $P(\psi,0)=\mathbf{n}(\psi),X(\psi,0)=0$, $\dot
P(\psi,0)=-C_x(0)=-\mathbf{C}'_0$,$\dot
X(\psi,0)=C_0\mathbf{n}(\psi)$ and
$\mathbf{m}(\Omega_j^{(1,0)})=\mathbf{m}(\Omega_j^{(0,2)})=0$ we
easily find in these maps
\begin{gather}
\nonumber K_{M^2_T}^h(f\mathbf{e}^0)|_{t=0}=\frac{i}{\sqrt{2\pi hC_0}}\times
\\ \nonumber\{
\int_{-\pi/4-\varsigma}^{\pi/4+\varsigma}\frac{e^{\frac{i((x_1 \cos \psi+x_2\sin
\psi)}{h}}}{|\cos \psi|}f(\rho,\psi) \exp\big({-i{\frac{x_2^2 \langle C_x(0),\mathbf{n}(\psi)\rangle}{2h C_0 \sin^2 \psi
}}}\big) \mathbf{e}_{j}(\psi)\mathbf{e}^0(x_1)\cos \psi d\psi-
\\\nonumber
\int_{3\pi/4-\varsigma}^{5\pi/4+\varsigma}\frac{e^{\frac{i((x_1 \cos \psi+x_2\sin
\psi)}{h}}}{|\cos \psi |}f(\rho,\psi) \exp\big({-i{\frac{x_2^2 \langle C_x(0),\mathbf{n}(\psi)\rangle}{2h C_0 \sin^2 \psi
}}}\big) \mathbf{e}_{j}(\psi)\mathbf{e}^0(x_1)\cos \psi d\psi+
\\
\nonumber
\int_{\pi/4-\varsigma}^{3\pi/4+\varsigma}\frac{e^{\frac{i((x_1 \cos \psi+x_2 \sin
\psi)}{h}} }{|\sin \psi |}f(\rho,\psi)\exp\big({i{\frac{x_1^2 \langle C_x(0),\mathbf{n}(\psi)\rangle}{2h C_0 \cos^2 \psi
}}}\big)
\mathbf{e}_{j}(\psi) \mathbf{e}^0(x_2) \sin \psi
d\psi-
\\\nonumber
\int_{5\pi/4-\varsigma}^{7\pi/4+\varsigma}\frac{e^{\frac{i((x_1 \cos \psi+x_2 \sin
\psi)}{h}} }{|\sin \psi |}f(\rho,\psi)\exp\big({i{\frac{x_1^2 \langle C_x(0),\mathbf{n}(\psi)\rangle}{2h C_0 \cos^2 \psi
}}}\big)
\mathbf{e}_{j}(\psi) \mathbf{e}^0(x_2) \sin \psi
d\psi\}.\end{gather}
Discarding  the absolute values under the square roots we see that sines and cosines in nominators and denominators are cancelled and only signs + appear before the integrals:
\begin{gather}\nonumber K_{\Lambda^2_t}^h(f\mathbf{e})|_{t=0}=\frac{i}{\sqrt{2\pi h C_0}}\times
\\ \nonumber\{
\int_{-\pi/4-\varsigma}^{\pi/4+\varsigma}{e^{\frac{i(x_1 \cos \psi+x_2\sin
\psi)}{h}}}f(\rho,\psi) \exp\big({-i{\frac{x_2^2 \langle C_x(0),\mathbf{n}(\psi)\rangle}{2h C_0 \sin^2 \psi
}}}\big) \mathbf{e}_{j}(\psi)\mathbf{e}(x_1)d\psi+\\
\nonumber
\int_{\pi/4-\varsigma}^{3\pi/4+\varsigma}{e^{\frac{i(x_1 \cos \psi+x_2 \sin
\psi)}{h}} }f(\rho,\psi)\exp\big({i{\frac{x_1^2 \langle C_x(0),\mathbf{n}(\psi)\rangle}{2h C_0 \cos^2 \psi
}}}\big)
\mathbf{e}_{j}(\psi) \mathbf{e}(x_2) d\psi+
\\
\nonumber
\int_{3\pi/4-\varsigma}^{5\pi/4+\varsigma}{e^{\frac{i(x_1 \cos \psi+x_2\sin
\psi)}{h}}}f(\rho,\psi) \exp\big({-i{\frac{x_2^2 \langle C_x(0),\mathbf{n}(\psi)\rangle}{2h C_0 \sin^2 \psi
}}}\big) \mathbf{e}_{j}(\psi)\mathbf{e}(x_1)d\psi+
\\
%\nonumber
\int_{5\pi/4-\varsigma}^{7\pi/4+\varsigma}{e^{\frac{i(x_1 \cos \psi+x_2 \sin
\psi)}{h}} }f(\rho,\psi)\exp\big({i{\frac{x_1^2 \langle C_x(0),\mathbf{n}(\psi)\rangle}{2h C_0 \cos^2 \psi
}}}\big)
\mathbf{e}_{j}(\psi) \mathbf{e}(x_2)
d\psi\}.\label{etaKO}   \end{gather}

Now let us put $h=\rho/\mu$ and  integrate this expression over $\rho$ from $0$ to $\infty$.
\begin{lemma} The following asymptotic equality is true:
$$\sqrt{\frac{ \mu C_0}{2\pi}}\rm{Re}\big( e^{-i\pi/4}\int_0^\infty d\rho\,\sqrt{\rho}\,K_{M^2_T}^{l/\rho}\tilde V(\rho \mathbf{n}(\psi) \mathbf{e})|_{t=0}\big)= V(x/\mu)+O(\mu).$$
\end{lemma}
\textsc{Sketch of Proof}. Let us first consider the last integrals without the factors  $\exp\big({-i{\frac{x_2^2 \langle C_x(0),\mathbf{n}(\psi)\rangle}{2h C_0 \sin^2 \psi
}}}\big) \mathbf{e}(x_1)$ and $\exp\big({i{\frac{x_1^2 \langle C_x(0),\mathbf{n}(\psi)\rangle}{2h C_0 \cos^2 \psi
}}}\big)\mathbf{e}(x_2)$. Then taking into account the equality $\sum_{j} \mathbf{e}_j=1$ we obtain  the integral
\begin{equation}\label{f1}
\frac{i}{\sqrt{2\pi \mu C_0}}
\int_{0}^{\infty}\int_{0}^{2\pi}{e^{\frac{i\rho(x_1 \cos \psi+x_2\sin
\psi)}{\mu}}}(\tilde V(\rho \mathbf{n}(\psi))\,\rho \,d\psi d\rho,
\end{equation}
which without the  factor $\frac{i}{\sqrt{\mu C_0 }}$ is no more but the inverse Fourier transform of the function $\tilde V(p)$. Thus this integral is  $V(x/\mu)$. Now we need to proof that the factors  $\exp\big({-i{\frac{x_2^2 \langle C_x(0),\mathbf{n}(\psi)\rangle}{2h C_0 \sin^2 \psi
}}}\big) \mathbf{e}(x_1)$ and $\exp\big({i{\frac{x_1^2 \langle C_x(0),\mathbf{n}(\psi)\rangle}{2h C_0 \cos^2 \psi
}}}\big)\mathbf{e}(x_2)$ change the function \eqref{f1} only by $O(\sqrt{\mu})$.
Consider for instance the expression corresponding to the first integral in \eqref{etaKO}
\begin{gather}\nonumber %\frac{i}{\sqrt{2\pi l C_0}}
\int_0^\infty \int_{-\pi/4-\varsigma}^{\pi/4+\varsigma}{e^{\frac{i\rho(x_1 \cos \psi+x_2\sin
\psi)}{l}}}\tilde V(\rho \mathbf{n}(\psi)\big(1- \exp\big({-i\rho {\frac{x_2^2 \langle C_x(0),\mathbf{n}(\psi)\rangle}{2l C_0 \sin^2 \psi
}}}\big)\big) \mathbf{e}_{j}(\psi)\rho d\psi d\rho
\end{gather}
We  divide this integral into two parts with the help of the cut off functions $\mathbf{e}_\rho(\rho/\sqrt{h})$ and  $1- \mathbf{e}_\rho(\rho/\sqrt{h})$. Here $\mathbf{e}_\rho(z)=1$ for $z<1/2$ and $\mathbf{e}_\rho(z)=0$ for $z>1$. It is simple to estimate the first part, which gives $O(h)$.  The second part we can integrate by part which gives the required estimate. The estimates of  other integrals are similar.

The verification of the second condition in $\eqref{WEq1}$ established by differentiation and similar calculations.
\begin{rem} The constructed solution could be decomposed into two parts. The first part is based on the used construction but with the different function $f$. The second part is similar, but based on characteristics and family of invariant  Lagrangian bands associated with the Hamiltonian $-|p|C(x)$. But all calculations related with this part is equivalent to complex conjugation of the second one, which  allows one not go beyond to this negative Hamiltonian.
\end{rem}

%Now let us show that the first derivative of .... on time $t$ for $t=0$ is equal to zero.

{\bf b.  Asymptotics corresponding to regular points.}
Now consider the neighborhood of the regular point. Then each  component from the first sum in \eqref{KO1} gives the following component in the solution \eqref{sol1}
\begin{gather} \nonumber
\frac{ \sqrt{2\pi l C_0}}{\sqrt{|\mathcal{J}_j^{(1,2)}(\psi,t)|}}\rm{Re}
\{ e^{-i\pi/4-\frac{i\pi}{2}\mathbf{m}(\Omega_j^{(1,2)})}
\int_0^\infty \sqrt{\rho}\,\,\tilde V(\rho \mathbf{n}(\psi))
\exp{\frac{i\rho S_{j}^{(1,2)}(x_1,x_2,t)}{\mu}}
d\rho \}\times\\\nonumber
\mathbf{e}_{j}(\psi)
\mathbf{e}^t(|x-X(\psi,t))|_{\psi=\psi^j(x_1,x_2,t)}.
\end{gather}
The Jacobian $|\mathcal{J}_j^{(1,2)}(\psi,t)|$ is $|\dot X||X_{\psi}|=C(X)|X_{\psi}|$, index $\mathbf{m}(\Omega_j^{(1,2)})=m(\psi^j(x,t),t)$ and the integral can be presented as a function $F(\frac{S_{j}^{(1,2)}(x_1,x_2,t)}{\mu},\psi^j(x,t))$.  The function $F(z,\psi)$ decays at least as $|z|^{-3}$ as $|z|\to \infty$. Thus the solution is localized in the neighborhood
of the front $\gamma_t$ and with the accuracy $O(\mu^{3/2})$ one can omit the cut off functions in this neighborhood. Also without disturbing of this accuracy one can change  $S_{j}^{(1,2)}(x_1,x_2,t)$ by its linear part $\langle P(\psi,t),(x-X(\psi,t))\rangle|_{\psi=\psi^j(x,t)}$: the quadratic correction to the phase  changes the solution
by  $O(\mu^{3/2})$ only. This gives the proof of Theorems 1 and 2.

{\bf c.  Asymptotics corresponding to singular maps.}
Finally let us consider briefly and without rigorous estimates the behavior of the solution in the neighborhood of a nondegenerated focal point. To this end we fix the time $t$ and the angle $\psi^F(t)$ determining the focal point $P^F=P(\psi^F(t),t),X^F=X(\psi^F(t),t)$ and rotated coordinates choosing the direction of the new axis $x'_2$ coinciding with the vector $X(\psi^F,t^F)$. It gives us opportunity to make all consideration in the map with $I=(0,2)$  and use formulas \eqref{coef}. Thus we need to investigate and to simplify  the integral
\begin{gather*}
e^{-\frac{i\pi}{2}\mathbf{m}(\Omega_j^{(0,2)})}
\frac{i}{\sqrt{2\pi h}}
\int_{-\infty}^{+\infty}\frac{\exp{\frac{i(S_{j}^{(0,2)}(p'_1,x'_2,t)+x'_1p'_1)}{h}}}
{\sqrt{|J_j^{(0,2)}(p'_1,x'_2)|}}
f(\rho,\psi)\mathbf{e}_{j}(\psi)|_{\psi=\psi_{j}(p'_1)}
\mathbf{e}^t(x'_2-X'_2) dp_1.
\end{gather*}
Like in the case for $t=0$ we change the integration over $dp'_1$ by the integration over $d\psi=dy$.  Taking into account formulas \eqref{change} we find:
\begin{gather*}
S_{j}^{(0,2)}(p'_1,x'_2,t)+x'_1p'_1|_{p'_1=P'_1}=\\P'_1(x'_1-X'_1)+P'_2(x'_2-X'_2)
+\frac{1}{2}(x'_2-X'_2(\psi,t))^2\frac{\dot P'_1 P'_{2\psi}-\dot P'_2P'_{1\psi}}{\dot P'_1 X'_{2\psi}-\dot X'_2 P'_{1\psi}}=\\ -q b(n+1)y x'_1+q an x'_2+q a b y^{n+1}+O(y^{n+2})+O(y^2 |x'|)
+O(x'_2-by^{n+1})^2
\end{gather*}
According to \eqref{deter} the Jacobian $J_j^{(0,2)}(p'_1,x'_2)=\tilde J^F+O(y)$.  We put $dP'_1=- q b(n+1)d y$, which gives in the integral the factor $|q b(n+1)|$ and the same limits $-\infty,\infty$.
It is possible to prove that in the neighborhood of the focal point (depended on $\mu$) after integration over $\rho$ the corrections denoted as $O(\cdot)$ gives the small correction to the leading term of asymptotic of $\eta$. Taking into account this fact and also the fact of fast decaying of  $\eta$  outside of the neighborhood of the front $\gamma_t$ we can omit the cut-off functions $\mathbf{e}_j$, $\mathbf{e}^t$. Finally we come to the following formula in the neighborhood of the focal point $X^F$
\begin{gather} \nonumber
\eta^F_j=\frac{|q b(n+1)| \sqrt{\mu C_0}}{\sqrt{2\pi|\tilde J_F|}}\times\\\rm{Re}
\{ e^{-i\pi/4-\frac{i\pi}{2}\mathbf{m}(\Omega_j^{(0,2)})}
\int_0^\infty d\rho\int_{-\infty}^\infty dy\rho\,\, \tilde V(\rho \mathbf{n}(\psi^F))\times\\
\exp{\frac{i\rho (-q b(n+1)y x'_1+q an x'_2+q a b y^{n+1})}{\mu}}\}
%\times\\\nonumber \mathbf{e}_{j}(\psi)\mathbf{e}^t(|x-X(\psi,t))|_{\psi=\psi^j(x_1,x_2,t)}.
\end{gather}
Now we express the coefficients $a,b,q$ via $J^{(n)},\tilde J^F, C_0,C_F$ using formulas \eqref{coef} and instead $y$ introduce the variable $\xi=\big|\frac{\tilde J_F J_F^{(n)}}{\mu C^2_F}\big|^{\frac{1}{n+1}}y$. After some algebra we obtain formula  \eqref{gn}.
\begin{rem} About the quadratic  terms (corrections) in the phases and regularization of asymptotic formulas. One can see that actually the quadratic terms disappear in the final formulas for the leading term of asymptotics. Nevertheless they play important role in our construction. First, without this term it is impossible to show that the asymptotic solution satisfy the original equation with necessary accuracy. Second the passage from representation acting in the regular maps to the representation acting in the singular maps is based on the partial  Fourier transform and 
stationary phase method 
(see \cite{ MaslovAsymptMethods,MaslovFedoryuk,DobrZh,Borovikov}). It is well known that the
application of  the second one based under assumption of nondegeneracy of second derivatives. Thus it is necessary to preserve the quadratic terms in proofs, and it is possible to omit the only in final formulas. It is also possible to proof that instead of the germ $\Lambda^2_t$ one can choose the germ  $\widetilde \Lambda^2_t$. This gives another asymptotic solution, but with the {\it same} leading term. More over this choice can be simplify the verification of the initial data because the exponential function containing
$x_1^2$ and  $x_2^2$  \eqref{etaKO} disappear (there are no quadratic with respect to  $x_1$ and  $x_2$ corrections in the phases) . Thus this choice can be viewed as the other regularization, but all of them are natural in such a sense that they based on {\it the more precise asymptotic construction}.
\end{rem}
 \section{Acknowledgments} This work was supported by RFBI- 05-01-00968.

\end{document}